\documentclass{article}

\usepackage{PRIMEarxiv}

\usepackage{silence}
\usepackage{graphicx}
\usepackage{booktabs}
\usepackage{multirow}
\usepackage{multicol}
\usepackage{threeparttable}
\usepackage{amsmath,amsfonts}
\usepackage[linesnumbered, 
            lined, 
            ruled, vlined]{algorithm2e}
\usepackage[caption=false,
            font=normalsize,
            labelfont=sf,textfont=sf]{subfig}
\usepackage[backend=biber,
            style=ieee,
            maxcitenames=2,
            mincitenames=1]{biblatex}
\usepackage[colorlinks=true, allcolors=blue]{hyperref}

\newcommand{\Aff}[1]{\textsuperscript{#1}}

\title{Attentive Dilated Convolution for Automatic Sleep Staging using Force-directed Layout%
\thanks{This work has been accepted for publication in IEEE Access. DOI: \href{https://doi.org/10.1109/ACCESS.2026.3680836}{10.1109/ACCESS.2026.3680836}}%
}

\author{
  Md Jobayer\Aff{1,2}\hspace{1pt}\thanks{Equal contribution.}\quad 
  Md Mehedi Hasan Shawon\Aff{1}\hspace{1pt}\footnotemark[2]\hspace{4pt}\thanks{Corresponding author: mehedi.shawon@bracu.ac.bd}\quad 
  Tasfin Mahmud\Aff{1,3} 
  \AND
  Md. Borhan Uddin Antor\Aff{1}\quad 
  Arshad M. Chowdhury\Aff{1} \\[4ex]
  \normalfont\small\Aff{1}Department of Electrical and Electronic Engineering, BRAC University, Bangladesh \\
  \small\Aff{2}Department of Biomedical Engineering, Linköping University, Sweden \\
  \small\Aff{3}Department of Electrical and Computer Engineering, Texas A\&M University, USA
}

\begin{document}

\maketitle

\begin{abstract}
Sleep stages play an important role in identifying sleep patterns and diagnosing sleep disorders. In this study, we present an automated sleep stage classifier called the Attentive Dilated Convolutional Neural Network (AttDiCNN), which uses deep learning methodologies to address challenges related to data heterogeneity, computational complexity, and reliable and automatic sleep staging. We employed a force-directed layout based on the visibility graph to capture the most significant information from the EEG signals, thereby representing the spatial-temporal features. The proposed network consists of three modules: the Localized Spatial Feature Extraction Network (LSFE), Spatio-Temporal-Temporal Long Retention Network (S2TLR), and Global Averaging Attention Network (G2A). The LSFE captures spatial information from sleep data, the S2TLR is designed to extract the most pertinent information in long-term contexts, and the G2A reduces computational overhead by aggregating information from the LSFE and S2TLR. We evaluated the performance of our model on three comprehensive and publicly accessible datasets, achieving state-of-the-art accuracies of 98.56\%, 99.66\%, and 99.08\% for the EDFX, HMC, and NCH datasets, respectively, while maintaining a low computational complexity with 1.4 M parameters. Our proposed architecture surpasses existing methodologies in several performance metrics, thereby proving its potential as an automated tool for clinical settings.
\end{abstract}


\section{Introduction}

Sleep is a fundamental physiological process that is vital for human health and well-being. It plays an important role in various aspects of life, including physiological health, cognitive function, emotional stability, and overall quality of life \cite{10.2147/nss.s163071}. The sleep process includes specific stages, each characterized by unique patterns of brain activity, eye movements, and muscle tone. These stages are categorized into two primary types: non-rapid eye movement (NREM) and rapid eye movement (REM). NREM is further divided into stages 1, 2, 3, and 4, each representing a progressively deeper level. Each stage fulfils specific functions, such as memory consolidation, hormonal regulation, and restoration of both the body and mind \cite{colten_sleep_2006}. Sleep duration and quality are vital determinants of overall health and homeostasis. Extensive research has demonstrated that poor sleep quality and inadequate sleep duration have deleterious health consequences, including cardiovascular disease, metabolic disorders, and impaired cognitive function \cite{10.3390/children8070542}. Furthermore, the role of the sleep cycle in facilitating post-stroke recovery and neurorehabilitation has been comprehensively investigated, underscoring the significance of sleep in neurological health \cite{duss_role_2017}. Irregularities or the absence of specific sleep stages are correlated with various sleep disorders. For instance, while healthy individuals go through NREM stages in their sleep, those with narcolepsy directly transition into REM sleep \cite{carskadon_chapter_2005}. Therefore, the correct identification and classification of these sleep stages are essential for understanding sleep patterns and diagnosing sleep disorders.

Polysomnography is one of the most widely used methods for analyzing sleep data \parencite{migovich_feasibility_2023}. This technique captures various physiological signals, including electrocardiograms (ECG) and electroencephalograms (EEG), thus providing a comprehensive perspective on the multifaceted dimensions of sleep \parencite{mirth_identification_2023, zhang_eeg_based_2018}. Nonetheless, this approach is prone to subjectivity errors and inconsistencies due to the large amount of data that need to be analyzed \parencite{migovich_feasibility_2023}. However, recently, there has been a paradigm shift towards automatic sleep stage classification through machine learning techniques with traditional and advanced deep learning frameworks \parencite{Zhu_Luo_Yu_2020}. Traditional machine learning methodologies generally require manual extraction of features, followed by classification using models such as multilayer perceptron, support vector machine (SVM), hidden Markov model, and Gaussian mixture model \parencite{zhu_ms_hnn_2023}. In contrast, deep learning techniques employ neural networks to autonomously learn features and classify sleep stages, potentially improving accuracy and efficiency compared with manual methods. A limitation of traditional methods is their reliance on manually extracted features, which may not represent all the important information in the data. Furthermore, traditional methodologies often struggle to recognize temporal patterns in longitudinal data, thus constraining their ability to accurately capture complex sleep patterns and stage transitions \parencite{liu_multi-scale_2022}. In addition, the precision of traditional algorithms in detecting different sleep stages tends to vary, thus requiring the adoption of more sophisticated techniques for increased performance \parencite{badiei_novel_2023}. Incorporating machine learning algorithms into the diagnosis of sleep disorders has significantly improved the capacity to manage large datasets and identify indicators of sleep-related anomalies \parencite{airlangga_evaluating_2024}. However, the effectiveness of machine learning models in classifying sleep stages depends highly on the quality and quantity of the data \parencite{lal_temporal_2023}. Furthermore, there is a risk of overfitting when the model is trained on limited datasets \parencite{joshi_deep_2021}. In addition, the substantial computational demands and resource requirements of certain machine learning techniques pose challenges for real-time applications and deployment in resource-constrained environments \parencite{wang_effective_2023}.

\begin{figure}
    \centering
    \includegraphics[width=\columnwidth]{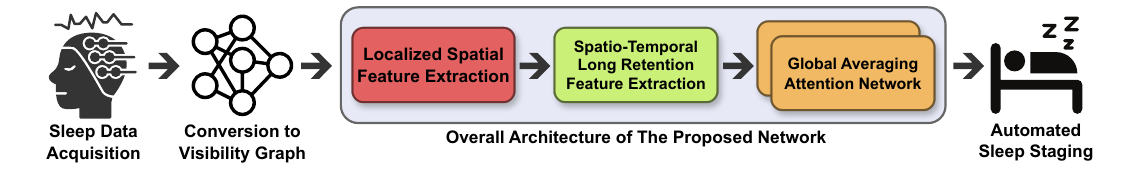}
    \caption{An overview of the proposed system. It begins with acquiring EEG data from the devices and converting them to visibility graphs. It is then passed to the model, and it captures all the relevant features of the data and predicts the sleep stage.}
    \label{fig_workflow}
\end{figure}

In this study, we introduce an automatic sleep stage classifier that utilizes deep learning algorithms to address the aforementioned limitations. The diagram in Figure \ref{fig_workflow} provides an overview of the proposed system. This study outlines the following major contributions to the field of automatic sleep stage classification:

\begin{itemize}
  \item We introduce a novel architecture, referred to as the Attentive Dilated Convolutional Neural Network (AttDiCNN), for automatic sleep staging. This architecture surpasses the performance of existing state-of-the-art models, demonstrating better performance and improved computational efficiency.
  \item Although there are existing studies on automatic sleep staging utilizing visibility graphs, to the best of our knowledge, this is the first attempt at employing Kamada-Kawai layout algorithms to produce a representation of the sleep data from visibility graphs and explore the potential for automatic sleep staging using this information.
  \item We developed two specialized modules to extract spatio-temporal details from the visibility graphs (VGs). To capture spatial data effectively, we used dilated receptive fields capable of detecting interdependencies among adjacent VG nodes at specific frequencies. Additionally, we applied a multihead attention network to the dilated data to capture the contextual-temporal details.
  \item To demonstrate the effectiveness of our model, we conducted an evaluation using three publicly available datasets, where we chose one of these datasets as a benchmark for comparative analysis with existing studies, given its extensive use in this domain.
\end{itemize}

\section{Related Work}

Various methodologies have been used to automatically detect sleep stages using traditional and advanced machine learning techniques. Conventional methods follow various techniques, such as the analysis of polysomnography (PSG) data, electrodermal activity (EDA), and sleep depth information derived from electroencephalography (EEG) data. For example, visual analysis of PSG data using multiple biosignals has proven to be an effective tool for manually evaluating sleep quality \parencite{najdi_feature_2017}. The PSG analysis process involves interpreting specific signal patterns and characteristics that adhere to standard guidelines. In contrast, the analysis of EDA patterns has been shown to identify specific characteristics correlated with sleep quality, thus facilitating the identification of sleep stages \parencite{herlan_electrodermal_2019, sano_quantitative_2014}. EDA reflects variations in skin conductance due to sweating-related changes, which correspond to different sleep stages, as the EDA pattern fluctuates with sleep stage transitions. In addition, estimating sleep depth from EEG data and identifying relevant sleep stages have shown promising potential. \textcite{kaplan_evaluation_2015} introduced the Z-PLUS algorithm, which estimates sleep depth and classifies it into different sleep epochs such as light sleep, deep sleep, and rapid eye movement. However, these traditional methodologies have several limitations: they are often time-consuming, subjective, and expensive to conduct \parencite{najdi_feature_2017}. Techniques involving visual analysis are particularly labor-intensive and susceptible to human error, leading to inconsistencies in sleep stage classification \parencite{imtiaz_systematic_2021}. Furthermore, they lack real-time detection capabilities, which require post-processing and manual review before concluding, making them less suitable for clinical environments \parencite{yildirim_deep_2019}. Traditional methods also struggle to capture the complexity of sleep dynamics, frequently overlooking subtle patterns and variations in sleep data \parencite{altini_promise_2021}.

In contrast, machine learning techniques have demonstrated their potential for automatic, rapid, and precise sleep staging processes. They are also known to handle large volumes of data and derive intricate and meaningful representations from them \parencite{ramachandran_survey_2021}. \textcite{10.1109/tbme.2018.2872652} introduced an innovative framework for automated sleep staging using a convolutional neural network (CNN), which simultaneously classifies sleep stages and predicts the epochs of adjacent labels. This method uses dependencies among consecutive epochs to enhance accuracy. Their proposed model is also capable of making multiple decisions by applying ensemble-of-model techniques, and the information is then aggregated to produce a final prediction. Although their ensemble methods outperformed singleton methods, they often resulted in computational overheads. Furthermore, the model's ability to generate multiple outputs based on probabilistic aggregation can diffuse the classification confidence, which is an undesirable trait in detection tasks. Similarly, \textcite{Emina_2018_Ensemble} proposed an ensemble-based classifier that combined principal component analysis and rotational SVM to classify the five sleep stages. However, their model, which relies on single-channel EEG data, does not exhibit a channel-agnostic behavior. Additionally, the use of nonlinear algorithms and principal component analysis simplifies the complex behavior of the data, leading to oversimplified decisions. \textcite{10.32604/csse.2023.030603} presented the Competitive Multi-Verse Optimization with a Deep Learning-based Stage Classification (CMVODL-SSC) model that uses EEG signals to classify sleep stages. This model incorporates data preprocessing and utilizes a cascaded long short-term memory (CLSTM) model, using the CMVO algorithm for hyperparameter tuning. Nevertheless, the use of LSTM models, which maintain long-term data, incurs memory overhead, and reliance on a single dataset restricts the model's generalization. Furthermore, \textcite{chriskos_automatic_2020} introduced a unique framework for automatic sleep staging using CNN and cortical connectivity image. However, cortical connectivity images have encountered several problems. For example, the utilization of a general head anatomy model limits adaptability to individual participants and decreases localization accuracy, necessitating expert intervention in the process. Moreover, the limited number of electrodes used to produce these images of cortical connectivity results in an ill-posed inverse problem owing to the mismatch between the number of electrodes and the numerous active sources in the cortex \parencite{filatova_dynamic_2018}.

In summary, an automated approach is required to accurately determine sleep stages with features that effectively extract all pertinent information from the signals and capture the complex dynamics of these features. Furthermore, the method should be computationally efficient for accurately classifying sleep stages across multiple datasets.

\begin{figure}[!ht]
    \centering
    \includegraphics[width=\columnwidth]{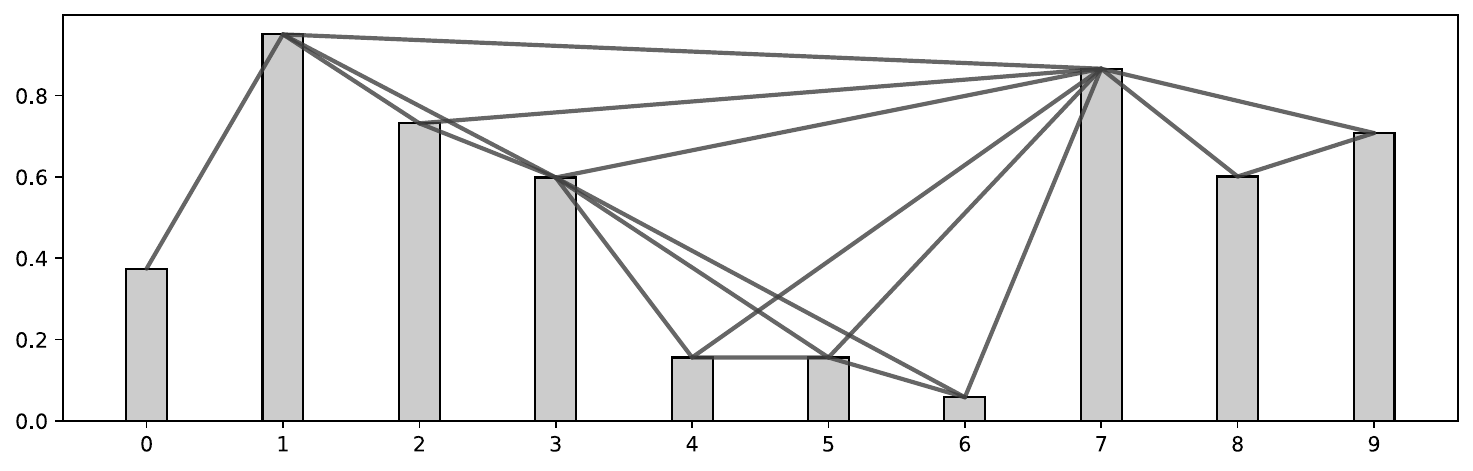}
    \caption{An undirected natural VG with 10 vertices equivalent to 10 series data in an ordered manner. The horizontal axis refers to the data points, and the vertical axis refers to their magnitudes. The connecting lines refer to the links between the nodes in the graph, indicating a relationship between them.}
    \label{fig_nvg_sample}
\end{figure}

\section{Materials and Methods}

\subsection{Construction of the Visibility Graph}

\begin{figure*}[!ht]
    \centering
    \includegraphics[width=\textwidth]{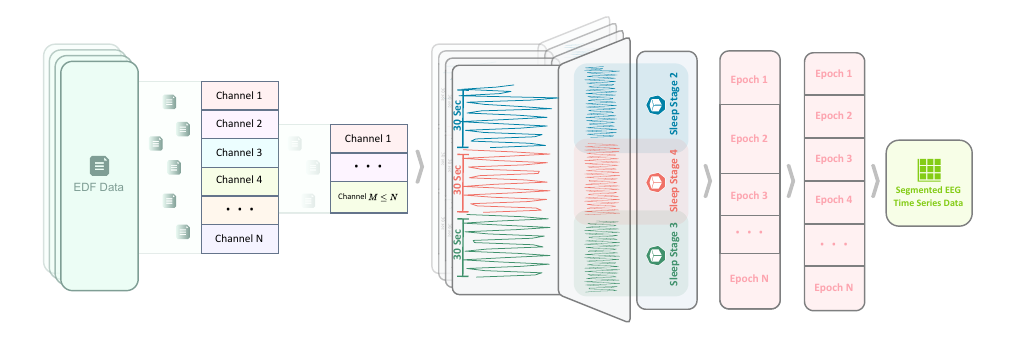}
    \caption{This illustration of the conversion process from the raw European Data Format (EDF) data into time-series data. The process can be divided into three sections: filtering the channels, mapping the epoch data with the corresponding sleep stage, and converting the data to time-series data. Channels are chosen based on the commonness of a channel across the datasets used in the experiment. The epochs have to be of the same length before the time-series conversion.}
    \label{fig_edf_conversion}
\end{figure*}

A VG is an algorithm that maps a series of data to a network graph \cite{lacasa_time_2008}. This graph provides an insightful visual representation of the time-series data, capturing the intrinsic properties of the series. Generally, there are two types of VG: natural and horizontal types. In this study, we focused on natural VG. Figure \ref{fig_nvg_sample} shows an undirected representation of the natural VG. In this graph, each vertex corresponds to a data point in the series that is arranged in the same sequence. The magnitudes of the vertices represent the actual values of the data points. The edges between the vertices are drawn based on the visibility between the vertices and their neighboring vertices. The overall process of converting the EEG data into series data is shown in Figure \ref{fig_edf_conversion}.

Consider $s(t_x)$ as a univariate time series, where $t_x$ denotes the time events defined as $\{0, 1, 2, \ldots N\}$ with $N$ being the total number of events. The construction of a VG involves creating $N$ vertices of the same order as $t_x$, with their magnitudes defined by the series values $s(t_x)$. Consider $t_i$ and $t_j$ as two random time events with corresponding vertices $s(t_i)$ and $s(t_j)$. The vertices are mutually visible if there exists a vertex $s(t_k)$ corresponding to the time event $t_k$ positioned between $s(t_i)$ and $s(t_j)$ such that $t_i < t_k < t_j$ and satisfies the following condition:
\begin{equation}\label{eq_nvg_visibility}
    s(t_k) < s(t_j) + [s(t_i) - s(t_j)]\dfrac{t_j - t_k}{t_j - t_i}
\end{equation}

In Equation \ref{eq_nvg_visibility}, an edge between $s(t_i)$ and $s(t_j)$ is established only if $s(t_k)$ does not intersect the visibility line extending from $s(t_i)$ to $s(t_j)$ or, equivalently, $s(t_k)$ resides below the line joining $s(t_i)$ and $s(t_j)$.

\subsection{Generating the Force-directed Layout}

\textcite{kamada_algorithm_1989} introduced the force-directed layout (FDL) as an undirected graph influenced by the spring model \cite{eades1984heuristic}. Consider a graph $G=\{V, E\}$ composed of a set of vertices $V$ and edges $E$. FDL aims to arrange the vertices $V$ such that the overall spring energy of the system is minimal, with the spring energy representing the graph-theoretic distance (i.e., Euclidean geometric distance) between each pair of vertices in the set $V$.

Consider a set of particles $P=\{P_1, P_2, P_3, \ldots, P_n\}$ corresponding to a series of vertices $V=\{V_1, V_2, V_3, \ldots, V_n\}$, where $n$ denotes the total number of particles. Each particle $P_{ij}$ is interconnected by springs. The total energy of the system, which is indicative of its degree of imbalance, $E$, can be formulated as follows:
\begin{equation}\label{eq_kamada_kawai_energy}
    E = \sum_{i=1}^{n-1} \sum_{j=i+1}^{n}\dfrac{1}{2}k_{ij}{(\mid P_i - P_j \mid - \: l_{ij})}^2
\end{equation}

\begin{figure*}[!htb]
    \centering
    \includegraphics[width=\textwidth]{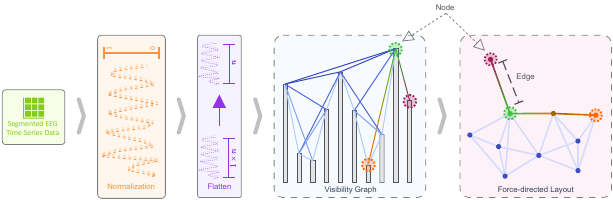}
    \caption{This illustration illustrates generating an FDL graph from time-series data. The conversion process can be divided into three steps: normalizing the data, constructing a visibility graph, and generating a force-directed layout. The FDL has been plotted utilizing the Kamada-Kawai layout algorithm.}
    \label{fig_eeg_to_fdl}
\end{figure*}

In Equation \ref{eq_kamada_kawai_energy}, $l_{ij}$ represents the minimum desired length of the spring that interconnects $p_i$ and $p_j$. It is mathematically expressed as $l_{ij} = L \times d_{ij}$, where $L$ denotes the preferred length of an individual edge within the display plane and $d_{ij}$ is the shortest path between the vertices $V_i$ and $V_j$. Furthermore, the parameter $k_{ij}$, indicative of the stiffness of the spring between $p_i$ and $p_j$, is defined as $k_{ij} = \frac{K}{d_{ij}^2}$, where $K$ is a constant. The method for creating an FDL from sleep data in the EDF is outlined in Algorithm \ref{algo_fdl_layout}, and an illustration is provided in Figure \ref{fig_eeg_to_fdl}.

\begin{algorithm}
\PrintSemicolon
\LinesNotNumbered
\SetAlgoLined\SetArgSty{}
\caption{Generation of Force-directed Layout}
\label{algo_fdl_layout}

\KwIn{Raw EEG data in \emph{EDF} format}
\KwOut{FDL graph images}
\KwData{EDF file paths as list}

$\mathcal{E} \gets \emptyset$\;

\ForEach{$(edfPath, annotPath)$ \textbf{in} $\exists paths$}{
	$eeg \gets readRawEdf(edfPath)$\;
        $eeg' \gets crop(eeg, [0, \exists t])$\;
        $eeg' \gets resample(eeg', \exists f)$\;
        $setAnnotation(eeg', annotPath)$\;
	$\mathcal{E} \gets \mathcal{E} + eeg$\;
}

$\mathcal{H} \gets \emptyset$\;

\ForEach{$eeg$ \textbf{in} $\mathcal{E}$}{
	$events \gets eventsFromAnnotation(eeg)$\;
        $epoch \gets Epoch(eeg, events, \ldots)$\;
	$\mathcal{H} \gets \mathcal{H} + epoch$\;
}

$\mathcal{D}(X, Y) \gets \emptyset: \{\mathbb{R}, \mathbb{N}\}$\;

\ForEach{$epoch$ \textbf{in} $\mathcal{H}$}{
	$x \gets getData(epoch)$\;
	$y \gets getEvent(epoch, [:, 2])$\;
	$\mathcal{D} \gets \mathcal{D} + (x, y)$\;
}

$\mathcal{I}(X, Y) \gets \emptyset: \{\mathbb{R}, \mathbb{N}\}$\;

\ForEach{$(X, Y)$ \textbf{in} $\mathcal{D}$}{
	$nvg = NaturalVG()$\;
        $vgTrain \gets nvg(X)$\;
        $img \gets drawFdl(vgTrain)$\;
	$\mathcal{I} \gets \mathcal{I} + (img, Y)$\;
}

\end{algorithm}

\subsection{Proposed Methodology}

\subsubsection{Convolutional Dilation}

A CNN is designed to adapt to multidimensional data invariances through local connection patterns with trainable kernels that are constrained by their weights. Standard CNNs consist of three primary layer types: convolutional, pooling, and fully connected layers. The convolutional layer generates the neuron output for local input regions by calculating the dot product of their weights and the current local region in which the convolution is applied. The pooling layer reduces the spatial dimensions of the output, thereby reducing the number of model parameters. The fully connected layer ultimately generates class scores via backpropagation. Dilated convolutions \parencite{yu_multi_scale_2015} are a crucial aspect of deep learning frameworks, allowing the expansion of the receptive field filter without increasing the parameter count or computational load \parencite{engelmann_dilated_2020}. This is achieved by incorporating gaps or holes between the elements of the convolutional kernel, effectively adding zero padding \parencite{schmidt_d2conv3d_2021}. By adding these holes, dilated convolutions enable the filters to gather information from a larger context while preserving the spatial resolution of the inputs. The dilation rate, which dictates the gaps between the kernel elements, controls the degree of expansion of the receptive field \parencite{ding_learning_2019}. By employing dilated convolutions, the models can effectively blend local and contextual information by capturing long-range dependencies and contextual details. This enhances their feature extraction capabilities across various scales and boosts their performance in tasks such as semantic segmentation and object detection.

\subsubsection{Multihead Self Attention}

An attention function is the process of mapping a query $Q$ and a set of key-value ($K-V$) pairs to an output, where the query, keys, values, and output are all vectors \cite{vaswani_2017_attention}. The output is computed as a weighted sum of the values, where the compatibility function of the query with the corresponding key computes the weight assigned to each value. Scaled dot-product attention is a member of the dot-product attention family, in which the attention score is calculated using the dot matrix multiplication of the input vectors. To avoid vanishing-gradient problems, the result of the dot product is scaled by a factor of $\sqrt{d_k}$ where $d_k$ is the dimension of the key vector. This method of scaling the dot product is called the scaled dot-product attention mechanism.

\begin{equation*}
    \text{Attention }(Q, K, V) = \text{softmax}(\dfrac{QK^T}{\sqrt{d_k}})V
\end{equation*}

\begin{figure}
    \centering
    \includegraphics[width=0.6\textwidth]{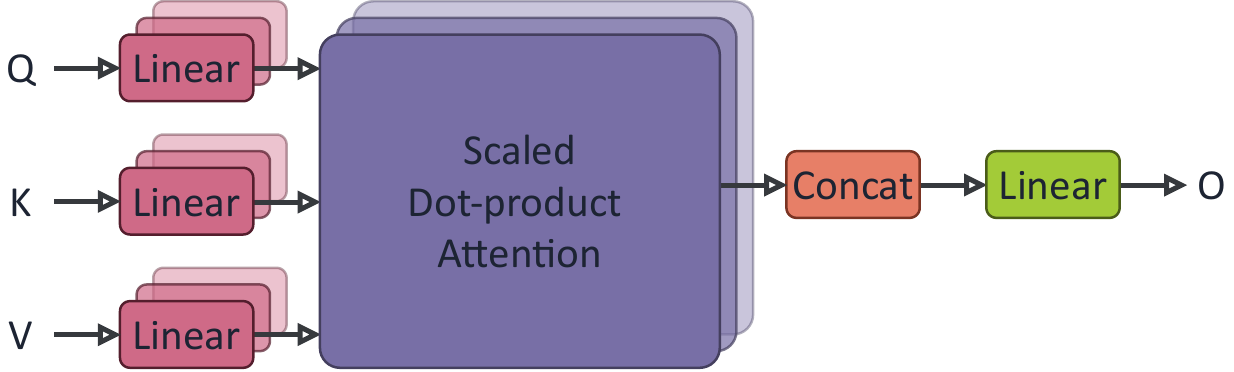}
    \caption{Multihead attention network composed of multiple attention layers operating concurrently. Each attention layer comprises a query vector $Q$, a key vector $K$, a value vector $V$, and an output vector $O$.}
    \label{fig_mha}
\end{figure}

Unlike single-head attention, multihead attention, as shown in Figure \ref{fig_mha}, linearly projects queries, keys, and values $h$ times with different learned linear projections to different dimensions. It transforms input queries, keys, and values into multiple subspaces using different learnable projection matrices. Each subspace is then used to compute the attention weights independently, allowing the model to focus on different aspects of the input. The outputs from these multiple attention heads are concatenated and transformed by a final linear layer to produce the final output.

\begin{equation}\label{eq_mha}
    \begin{aligned}
        \text{Multihead}(Q, K, V) &= \text{concat}(\text{head}_1, \ldots, \text{head}_h) W^o \\
        \text{where head}_i &= \text{Att}(QW_i^Q, KW_i^K, VW_i^V) \\
        \text{where Att}(Q, K, V) &= \text{Attention}(Q, K, V)
    \end{aligned}
\end{equation}

In this context, the projections are delineated by parameter matrices as follows:
\begin{equation*}
    \begin{aligned}
        W_i^Q &\in \mathbb{R}^{d_{model} \times d_k} & W_i^K &\in \mathbb{R}^{d_{model} \times d_k} \\
        W_i^V &\in \mathbb{R}^{d_{model} \times d_v} & W^O &\in \mathbb{R}^{hd_v \times d_{model}} \\
    \end{aligned}
\end{equation*}

\subsubsection{Network Architecture}

\begin{figure*}[!htb]
    \includegraphics[width=\textwidth]{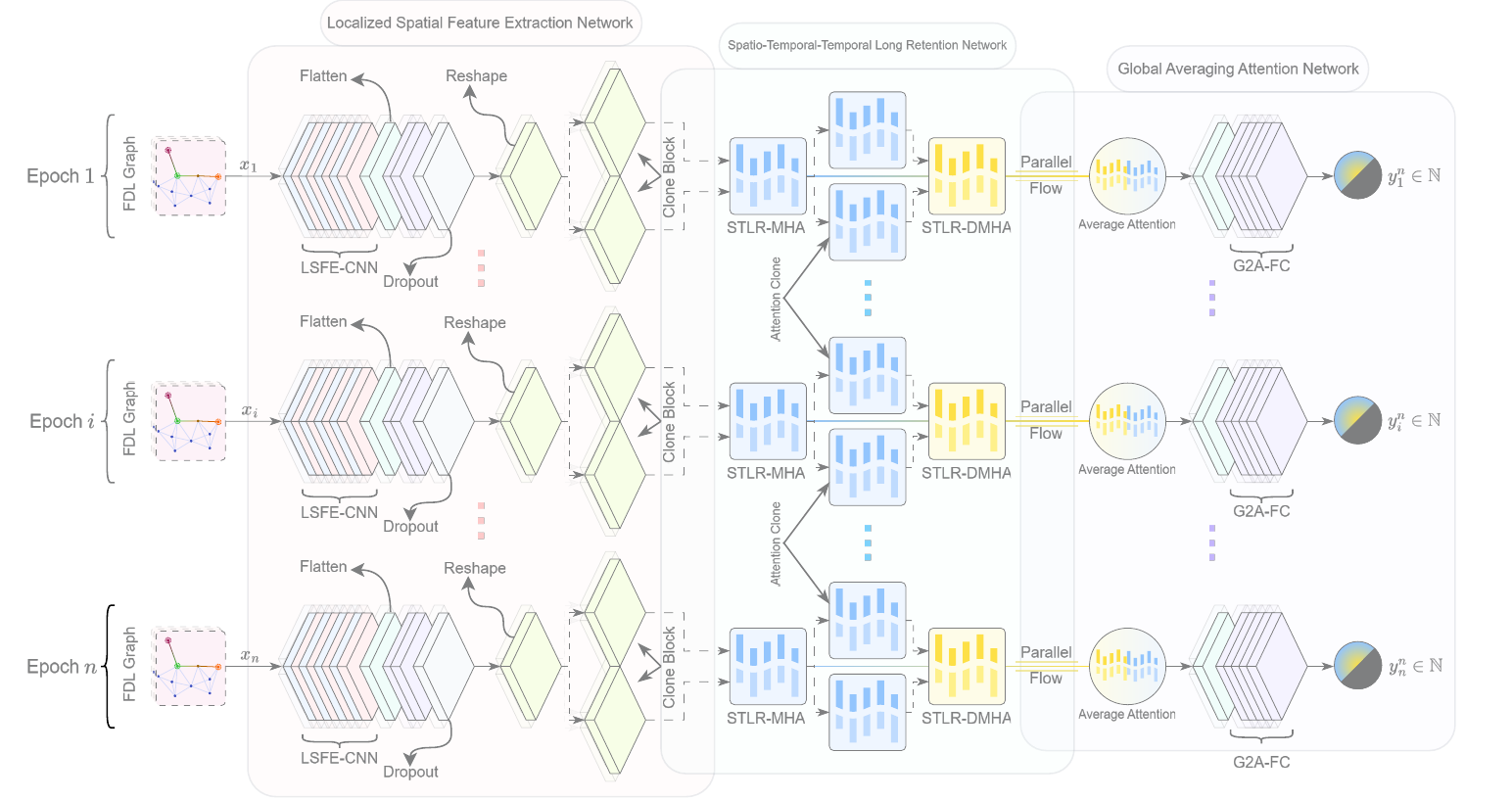}
    \caption{The illustration of the proposed model architecture. The network comprises three distinct blocks: LSFE, S2TLR, and G2A. LSFE extracts information based on local features, whereas S2TLR extracts meaningful and computationally efficient data from the retained long information stack. Finally, G2A averages the attention weights and concludes with a sleep stage.}
    \label{fig_mdl_arch}
\end{figure*}

We present our proposed architecture, as shown in Figure \ref{fig_mdl_arch}, which is designed for the automatic classification of sleep stages. This comprehensive network encompasses three different modules: the Localized Spatial Feature Extraction Network (LSFE), the Spatio-Temporal-Temporal Long Retention Network (S2TLR), and Global Averaging Attention Network (G2A). Each module generates a unique set of feature maps denoted by $\mathcal{F}_n$, where $n$ represents the total number of feature maps in the module. The following sections provide an in-depth discussion of each of these modules.

\paragraph{Localized Spatial Feature Extraction Network (LSFE)}

The LSFE is the first module of the proposed architecture, which captures the local characteristics of the FDL images. Let $\mathcal{I}$ represent the training set of size $S$, defined as $\{\mathcal{I}_s\}^S_{s=1}$. Each input, $\mathcal{I}_i$, is characterized as $\mathcal{I}_i = \{x_i, y_i\}$, where $x_i$ denotes the FDL image tensor of dimensions $(128, 128)$, that is, $x = \{x_{128} \times x_{128}\}$. The variable $y_i$ corresponds to the annotated sleep stage of the respective FDL data $x_i$. To encapsulate the local features of the FDL image, we developed a structure designated as \textit{LSFE-CNN}, which is composed of two standard CNN layers, two dilated CNN layers, a single flatten layer, and two fully connected layers. The aggregation of these layers forms the initial feature map $\mathcal{F}_1$, which encapsulates all the local variations in the data. Four maximum pooling layers succeeded the four CNNs. For the standard CNNs, each possesses a configuration $\{H, W, C\}$ where $(H, W)\equiv(2,2)$ indicates the kernel size, and $C$ denotes the number of kernels, which are 16 and 32 for the first two CNNs, respectively. Conventional CNNs traverse the entire FDL graph with a receptive field window of size $[2,2]$ and extract the most pertinent information using a maximum-pool strategy, determined by the formula: $max(0, p)$ where $p$ represents the scanning data point. In contrast, for the dilated CNNs, each is configured with $\{H, W, D, C\}$ where $(H, W)\equiv(2, 2)$ is the kernel size, $D=2$ is the dilation rate, and $C = 64 \text{ and } 128$ are the number of kernels, respectively. Similar to conventional CNNs, dilated CNNs also employ a maximum-pooling strategy. However, unlike conventional methods, the size of the receptive field can be adjusted by a factor of $l$, as delineated below:
\begin{equation*}
    (F \ast \! _{l} \; k) (p) = \sum_{s+ lt = p} F(s) \; k(t)
\end{equation*}

In this study, we consider $F : \mathbb{Z}^2 \rightarrow \mathbb{R}$ as a discrete function and $k: \Omega_r \rightarrow \mathbb{R}$ as a discrete filter size, where $\Omega_r = [-r, r]^2 \cap \mathbb{Z}^2$ \cite{yu_multi_scale_2015}. The proposed architecture employs a dilation rate of $l=2$, effectively skipping one data point along each of the three dimensions. This helps capture a long range of data with computational requirements comparable to those of conventional CNNs, thereby expediting the detection of salient features. Following each CNN layer, the rectified linear unit (ReLU) activation function was used to address the vanishing gradient problem. Subsequently, two fully connected networks were deployed, comprising 256 and 128 neural units, respectively, followed by a flatten layer. To prevent overfitting, a dropout layer with a rate of 0.5 was incorporated into the final layer of this network.

\paragraph{Spatio-Temporal-Temporal Long Retention Network (S2TLR)}

The purpose of this module is to encapsulate the most relevant information within an extended temporal event-data structure. The feature map $\mathcal{F}_2$ consists of two instances of reshaped local features derived from the LSFE network, which are sequentially processed through the multihead attention (MHA) network. Through convolution via the LSFE network, we obtained the most relevant spatial features, which were then fed into the S2TLR network. Let $\mathcal{I}_{S2TLR}$ represent the input data to the S2TLR network, characterized by dimensions $(b, 1, 128)$, where $b$ is the batch size. In our experiments, we utilized batch sizes of $b = \{32, 64, 128, 256, 512, 1024\}$. The input data $\mathcal{I}_{ST}$ are configured as $(\mathcal{I}_{ST}, \mathcal{I}_{ST})$ and fed into the MHA with $h = 3$ heads to compute the temporal relationships among the spatial features, referred to as self-attention weights. We denote this attention as $\mathcal{A}_{ST} = Att(\mathcal{I}_{ST}, \mathcal{I}_{ST})$, as shown in Equation \ref{eq_mha}. This configuration, designated as S2TLR-MHA, implies the spatio-temporal interaction of the convolved features from the LSFE network. The $\mathcal{A}_{ST}$ output is subsequently embedded in another MHA network, producing data referred to as $\mathcal{I}_{TT}$, which symbolizes the temporal-temporal relationship of the spatio-temporal features. The data $\mathcal{I}_{TT}$ are processed in the configuration $(\mathcal{I}_{TT}, \mathcal{I}_{TT})$ through the network with the same number of heads as $\mathcal{A}_{ST}$, and this attention is referred to as $\mathcal{A}_{TT} = Att(\mathcal{I}_{TT}, \mathcal{I}_{TT})$. This represents the S2TLR-DMHA block depicted in Figure \ref{fig_mdl_arch}, thereby completing the spatial-temporal-temporal computational pipeline.

\paragraph{Global Averaging Attention Network (G2A)}

The G2A network balances the localized weights $W_l$ and global attention weights $W_g$, which represent the feature map $\mathcal{F}_3$. This network reduces the computational overhead by half by averaging both the local and global weights. This approach facilitates the optimization and equilibrium of the resultant attention. The optimized average is expressed as follows:

\begin{equation*}
    W_o = Avg(W_l, W_g) = \dfrac{W_l^{(1, 128)} + W_g^{(1, 128)}}{2}
\end{equation*}

In this context, $W_o$ represents the optimized weight, maintaining the identical data dimensionality as $W_l$ and $W_g$, specifically, $(1, 128)$. The local weight $W_l$ corresponds to the attention weights from the $\mathcal{A}_{ST}$ block, which encompasses the S2TLR-MHA features, whereas the global weight $W_g$ pertains to the attention weights from the $\mathcal{A}_{TT}$ block, which incorporates the S2TLR-DMHA features. $W_o$ enables a balanced attention mechanism by combining the changes in the spatial dimension captured by the S2TLR-MHA and the temporal retention of these features of the S2TLR-MHA throughout an entire epoch, as captured by the S2TLR-DMHA. Subsequently, $W_o$ is flattened and passed through a sequence of four fully connected (FC) layers with neural units of $\{512, 128, 64, 32\}$, respectively, to yield the final output, which consists of scoring the targeted sleep stage. This is accomplished by channeling the data through a final FC layer with a neural unit of $n$, which denotes the number of sleep stages in the specific dataset.

\section{Experimental Setup}

\subsection{Dataset Description}

We used three publicly available datasets to evaluate the performance of the proposed model. One of these datasets is frequently utilized in the existing literature; therefore, we selected it as the benchmark dataset in our study. The remaining datasets were used to assess the robustness of our model because of their varying signal properties. The details of the three datasets are provided in the subsequent sections, and a summary of the datasets is presented in Table \ref{tab_ds_description}.

\subsubsection{EDFX}

The Sleep EDFX dataset \parencite{867928} from PhysioNet comprises two subsets: Sleep Cassette (SC) and Sleep Telemetry (ST), with a total of 197 whole-night PSG recordings, including EEG, electrooculogram (EOG), electromyography (EMG), and event markers. The first subset studied the impact of aging on sleep and included 153 SC files of patients aged 25--101 years. The second subset analyzed the effects of temazepam on sleep and contained 44 ST files from 22 unique subjects. The EOG and EEG signals were acquired at a sampling frequency of 100 Hz. In our study, we selected the Fpz-Cz channel to facilitate the automated sleep analysis process. The PSG files encompass the whole-night's sleep recordings, while the hypnogram files provide annotations of sleep patterns including stages W, R, 1, 2, 3, 4, M (movement time), and ? (not scored). Among these eight classifications, stage M was excluded from our study. The remaining sleep stages were included, and their distribution is shown in Supplementary Figure \ref{fig_class_dist}(a).

\begin{table}[!htb]
\centering
\caption{Properties of the three used datasets.}
\label{tab_ds_description}
\begin{tabular}{@{}llllll@{}}
\toprule
Ref & Dataset & EEG Channel & \#Class & Sampling Rate & \#Sample \\ \midrule
\cite{867928} & EDFX & Fpz-Cz & 7 & \multirow{3}{*}{100 Hz} & \multirow{3}{*}{25000} \\
\cite{lee_nch_nodate,lee_large_2022} & HMC & \multirow{2}{*}{C3-M2} & 5 &  &  \\
\cite{alvarez-estevez_inter-database_2021, alvarez-estevez_haaglanden_nodate} & NCH &  & 6 &  &  \\ \bottomrule
\end{tabular}
\end{table}

\subsubsection{HMC}

The Haaglanden Medisch Centrum sleep staging dataset \parencite{alvarez-estevez_inter-database_2021, alvarez-estevez_haaglanden_nodate} consists of 151 full-night PSG recordings collected from a diverse group of individuals referred for PSG examination at the Haaglanden Medisch Centrum (HMC, The Netherlands) Sleep Center. The dataset comprised recordings from 85 male and 66 female patients with an average age of 53.9 ± 15.4 years, covering a variety of sleep disorders. The PSG recordings included EEG, EOG, chin EMG, and ECG activity, in addition to event annotations for sleep stage scoring performed by HMC sleep annotators. The PSG data included EEG from four channels: F4-M1, C4-M1, O2-M1, and C3-M2, along with other EOG, EMG, and ECG channel data. However, in our study, we focused only on the C3-M2 channel for sleep analysis. Although originally recorded at 256 Hz, the data were resampled to 100 Hz for consistency. The class distribution for each stage in the HMC dataset is shown in Supplementary Figure \ref{fig_class_dist}(b).

\subsubsection{NCH}

The Sleep DataBank from the Nationwide Children's Hospital (NCH) \parencite{lee_nch_nodate, lee_large_2022} is an extensive dataset comprising 3,984 pediatric sleep studies performed on 3,673 distinct patients at the NCH in the USA. This dataset contains longitudinal clinical data sourced from the Electronic Health Record (EHR), which includes information on encounters, medications, measurements, diagnoses, and procedures, as well as published polysomnography (PSG) data with physiological signals. The unique aspects of this dataset are its significant size, specific focus on pediatric patients, real-world clinical environment, and comprehensive clinical data. We used the C3-M2 EEG channel to classify sleep stages. Supplementary Figure \ref{fig_class_dist}(c) shows the distribution of sleep stages within the NCH databank.

\subsection{Handling Data Imbalance}

Datasets frequently demonstrate notable class imbalances, causing overfitting and bias towards the prevalent class. To address this issue, we can either downsample or oversample the data. Downsampling may not be effective if other classes have too few samples, as the model would lack sufficient data to learn from. Thus, using the oversampling technique is more meaningful for generating synthetic data that resemble the original data. In our study, we used the synthetic minority sampling technique (SMOTE) \cite{chawla_smote_2011} to handle the imbalanced datasets. This technique generates synthetic data for minority classes to balance the data distribution. SMOTE creates new synthetic data by interpolating between existing instances of minority classes. It selects a minority sample, identifies its k nearest neighbors in the feature space based on the oversampling requirement, and creates a new synthetic sample by randomly choosing one of these neighbors.

\subsection{Training Strategy}

To achieve optimal model performance, it is essential to accurately balance the data and strategically divide them for training and testing. In our experiments, we used a random seed value of 13 to ensure the reproducibility of the results. The model training involved 10-fold stratified cross-validation across all datasets. We split the datasets into training and testing sets with an 80:20 ratio to evaluate the model performance on new, unseen data. The model was trained using sparse categorical cross-entropy as the loss function. Training was performed over 200 epochs using the Adam Optimizer with a learning rate of 0.001. During training, an early stopping mechanism was used with a patience threshold of 15, monitoring the validation accuracy in maximum mode. The overall process of training the proposed network and automatic sleep staging is described in Algorithm \ref{algo_proposed_workflow}.

\begin{algorithm}
\PrintSemicolon
\LinesNotNumbered
\SetAlgoLined\SetArgSty{}
\caption{Proposed Workflow}
\label{algo_proposed_workflow}

\KwIn{Force-directed Layout data}
\KwOut{Model's prediction}
\KwData{Network graphs as image files}

$\mathcal{I} \gets \emptyset$ \tcp*[f]{empty list}\;

\ForEach{$path$ \textbf{in} $paths$}{
	$img \gets loadImgAsArray(path)$\;
	$\mathcal{I} \gets \mathcal{I} \mid\mid img$ \tcp*[f]{append to the list}\;
}

$labelPath \gets \; ''\ldots label.csv''$\;
$\mathcal{Y} \gets loadLabelsFromCsv(labelPath) \in \mathbb{W}$\;

$\mathcal{I} \gets N(\mathcal{I})$ \tcp*[f]{normalize}\;
$(\tilde{\mathcal{I}}, \tilde{\mathcal{Y}}) \gets smote(\mathcal{I}, \mathcal{Y})$ \;

$k \gets 10$ \;
$\mathcal{F} \gets skf(k, \tilde{\mathcal{I}}, \tilde{\mathcal{Y}})$ \tcp*[f]{stratified k-fold} \;

$\mathcal{M} \gets AttDiCNN(\ldots)$\;
$modelTrain(\mathcal{M}, \mathcal{F}, \ldots, \tilde{\mathcal{I}}, \tilde{\mathcal{Y}})$\;

$\mathcal{P} \gets \mathcal{M}(\mathcal{I}) \in \mathbb{W}$ \tcp*[f]{model's prediction}\;

\end{algorithm}

\section{Results and Discussions}

\subsection{Evaluation Metrics}

A comprehensive set of evaluation metrics was used to assess the performance of our model. The metrics include accuracy (Acc), top-k accuracy, Cohen's kappa coefficient $\kappa$ \parencite{cohen_coefficient_1960}, area under the curve (AUC), precision, recall, and macro F1 score. Additionally, we used two error metrics, mean absolute error (MAE) and mean squared error (MSE), to record the model's error rate in predicting sleep stages.

\subsection{Performance Evaluation}

We used two different approaches: intra performance and inter performance comparisons to evaluate the performance of our proposed model. In the case of intra performance, we tested the performance variations across various iterations, whereas in the case of inter performance, we compared the performance of our model with the existing literature. The results of these two settings are described in the following sections.

\subsubsection{Intra Performance}

In this section, we evaluate the performance of the proposed model for different batch sizes to observe how the model behaves with varying amounts of data processed simultaneously. The batch sizes ranged from 32 to 1024 for each dataset, and the performance of the model varied with these changes, as detailed below.

\begin{table*}[!htb]
\centering
\caption{Model performance with different parameters.}
\label{tab_mdl_perf}
\resizebox{\textwidth}{!}{%
\begin{tabular}{@{}cccccccccccc@{}}
\toprule
\multirow{2}{*}{Dataset} & \multirow{2}{*}{Batch Size} & \multicolumn{10}{c}{Parameters} \\ \cmidrule(l){3-12} 
 &  & Acc. & Top-2 Acc. & Top-3 Acc. & Kappa $\kappa$ & AUC & Precision & Recall & MF1 & MAE & MSE \\ \midrule
\multirow{6}{*}{EDFX} & 32 & 0.8456 & 0.9780 & 0.9928 & 0.8043 & 0.9401 & 0.8995 & 0.9095 & 0.9028 & 0.2700 & 0.8995 \\
 & 64 & 0.9184 & 0.9866 & 0.9950 & 0.8919 & 0.9666 & 0.9181 & 0.9491 & 0.9316 & 0.1460 & 0.6239 \\
 & 128 & 0.9708 & 0.9966 & 0.9982 & 0.9629 & 0.9884 & 0.9760 & 0.9822 & 0.9787 & 0.0528 & 0.3883 \\
 & 256 & 0.9778 & 0.9964 & 0.9976 & 0.9718 & 0.9892 & 0.9793 & 0.9824 & 0.9807 & 0.0626 & 0.4917 \\
 & 512 & 0.9712 & 0.9952 & 0.9980 & 0.9635 & 0.9892 & 0.9754 & 0.9835 & 0.9793 & 0.0594 & 0.4443 \\
 & 1024 & \textit{0.9856} & \textit{0.9982} & \textit{0.9992} & \textit{0.9817} & \textit{0.9932} & \textit{0.9863} & \textit{0.9890} & \textit{0.9877} & \textit{0.0354} & \textit{0.3541} \\ \midrule
\multirow{6}{*}{HMC} & 32 & 0.3866 & 0.5706 & 0.7282 & 0.0415 & 0.5171 & 0.7248 & 0.2272 & 0.1629 & 1.0952 & 1.5436 \\
 & 64 & 0.5900 & 0.7370 & 0.8440 & 0.4009 & 0.6787 & 0.8405 & 0.4852 & 0.5417 & 0.7166 & 1.2467 \\
 & 128 & 0.9940 & 0.9988 & \textit{\textbf{1.0000}} & 0.9921 & 0.9971 & 0.9930 & 0.9956 & 0.9943 & 0.0142 & 0.1934 \\
 & 256 & 0.9892 & 0.9994 & \textit{\textbf{1.0000}} & 0.9858 & 0.9927 & 0.9872 & 0.9880 & 0.9875 & 0.0212 & 0.2154 \\
 & 512 & 0.9918 & \textit{\textbf{0.9998}} & \textit{\textbf{1.0000}} & 0.9892 & 0.9947 & 0.9917 & 0.9915 & 0.9915 & 0.0120 & \textit{0.1470} \\
 & 1024 & \textit{\textbf{0.9966}} & \textit{\textbf{0.9998}} & \textit{\textbf{1.0000}} & \textit{\textbf{0.9955}} & \textit{\textbf{0.9977}} & \textit{\textbf{0.9960}} & \textit{\textbf{0.9962}} & \textit{\textbf{0.9961}} & \textit{\textbf{0.0086}} & 0.1556 \\ \midrule
\multirow{6}{*}{NCH} & 32 & 0.9780 & 0.9972 & \textit{0.9996} & 0.9671 & 0.9904 & 0.9860 & 0.9859 & 0.9857 & 0.0248 & 0.1744 \\
 & 64 & 0.9702 & 0.9964 & \textit{0.9996} & 0.9556 & 0.9880 & 0.9678 & 0.9824 & 0.9744 & 0.0356 & 0.2209 \\
 & 128 & \textit{0.9908} & 0.9972 & \textit{0.9996} & \textit{0.9862} & \textit{0.9947} & \textit{0.9939} & \textit{0.9914} & \textit{0.9926} & \textit{0.0136} & 0.1497 \\
 & 256 & 0.9732 & 0.9966 & \textit{0.9996} & 0.9599 & 0.9888 & 0.9835 & 0.9838 & 0.9833 & 0.0324 & 0.2088 \\
 & 512 & 0.9866 & \textit{0.9974} & \textit{0.9996} & 0.9799 & 0.9933 & 0.9913 & 0.9896 & 0.9903 & 0.0160 & \textit{\textbf{0.1456}} \\
 & 1024 & 0.9794 & 0.9972 & \textit{0.9996} & 0.9692 & 0.9908 & 0.9877 & 0.9865 & 0.9870 & 0.0284 & 0.2098 \\ \midrule
\multicolumn{2}{c}{Minimum} & \textbf{0.3866} & \textbf{0.5706} & \textbf{0.7282} & \textbf{0.0415} & \textbf{0.5171} & \textbf{0.7248} & \textbf{0.2272} & \textbf{0.1629} & \textbf{0.0086} & \textbf{0.1456} \\ \midrule
\multicolumn{2}{c}{Maximum} & \textbf{0.9966} & \textbf{0.9998} & \textbf{1.0000} & \textbf{0.9955} & \textbf{0.9977} & \textbf{0.9960} & \textbf{0.9962} & \textbf{0.9961} & \textbf{1.0952} & \textbf{0.2209} \\ \midrule
\multicolumn{2}{c}{Average} & \textbf{0.9164} & \textbf{0.9577} & \textbf{0.9750} & \textbf{0.8777} & \textbf{0.9439} & \textbf{0.9543} & \textbf{0.9111} & \textbf{0.9082} & \textbf{0.1469} & \textbf{0.4340} \\ \bottomrule
\end{tabular}%
}
\end{table*}

\begin{figure*}[!htb]
    \includegraphics[width=\textwidth]{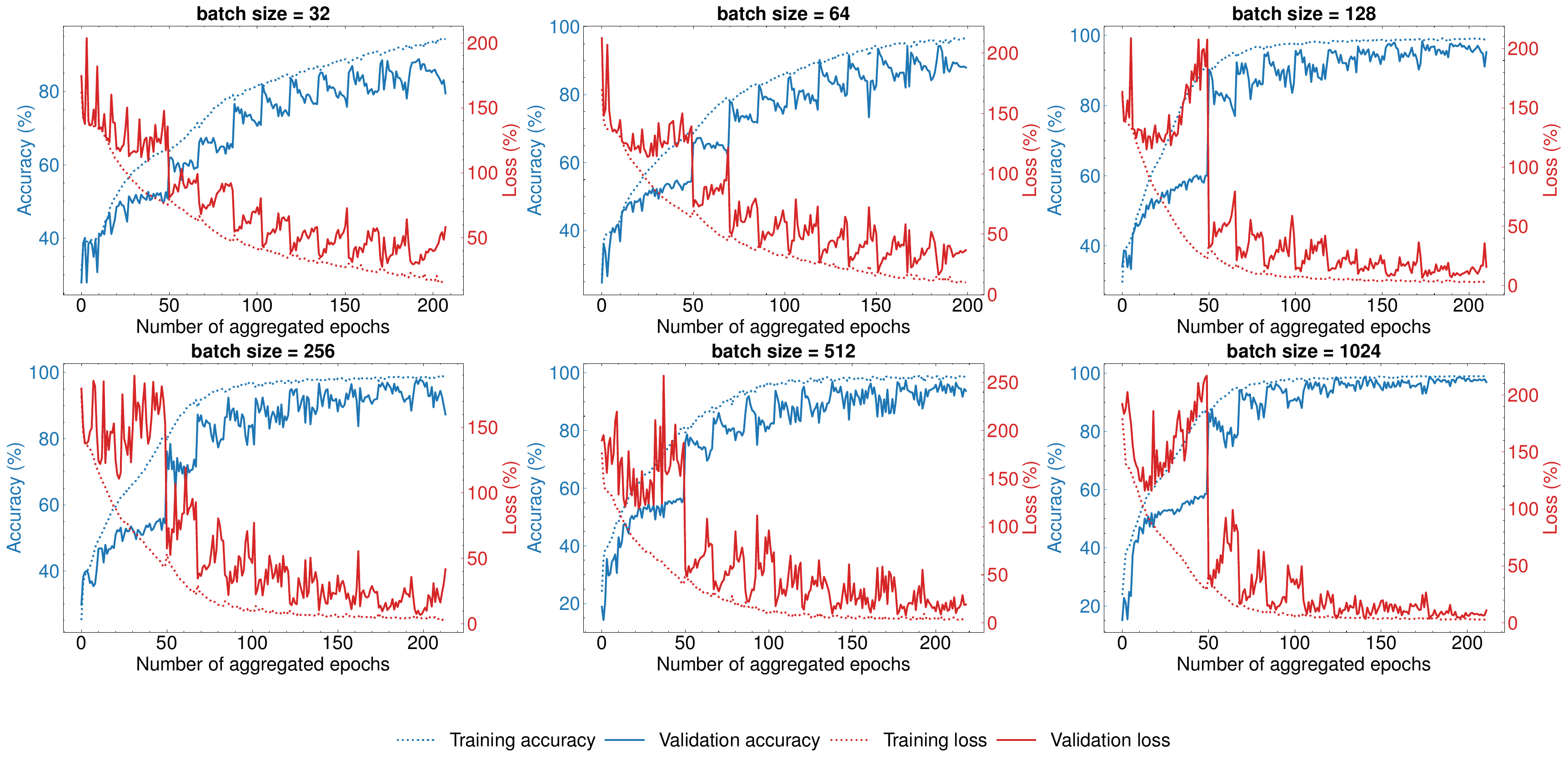}
    \caption{Performance metrics (accuracy and loss) plotted over epochs during training and validation for the EDFX dataset, using batch sizes of 32, 64, 128, 256, 512, and 1024.}
    \label{fig_edfx_train_val_curve}
\end{figure*}

As shown in Table \ref{tab_mdl_perf}, in the case of the EDFX dataset, we have a minimum accuracy of 84.56\% and a maximum accuracy of 98.56\%. This implies that our proposed model correctly predicted the sleep stages at least 84.56\% of the time in all instances. This also means that our model could not correctly predict 1.44\% of the time, even when it performed the best in classifying the sleep stages. However, the difference in the incorrect classification of stages is reduced in the case of top-2 and top-3 accuracies, with maximum scores of 99.82\% and 99.92\%, respectively. This implies that even if the model failed to predict in the first instance, it predicted almost perfectly almost all the time at its second and third instances. In Figure \ref{fig_edfx_train_val_curve}, it is evident that the model continued to improve its prediction capability as the number of epochs increased. The model appeared to learn smoothly during the training phase. However, it struggled to predict the validation set and had a scattered outcome, which slowly submerged and became very smooth at the end. In the case of error reduction, the MAE remained smooth almost all the time and improved slowly as the batch size increased. However, the MSE appears to be very large in smaller batch sizes and improves with time. The large MSE values are due to their higher sensitivity because of the squared nature of the data. Our proposed model had an MAE of 0.0354 and MSE of 0.3541 for the largest batch size of 1024.

Table \ref{tab_mdl_perf} shows a clear contrast in the performance between the EDFX and HMC datasets, particularly in the initial stages. The HMC dataset performed poorly with a batch size of 32. Initially, it had an accuracy of only 38.66\%. However, it converged faster starting from a batch size of 128, and was slightly higher than the performance of the EDFX dataset for the same batch size. Soon after, it outperformed the EDFX model. The superiority is also noticeable in the case of top-2 and top-3 accuracies, particularly for the top-k performance, which was predicted ideally starting from a batch size of 128 onward. There was also a visible difference in the reduction of false predictions, that is, the model's error rate. The HMC model performed very well in terms of the error metrics for the same batch size compared to the EDFX dataset. For the HMC dataset, we obtained minimum errors of 0.0086 and 0.1556 for the MAE and MSE, respectively. As shown in Supplementary Figure \ref{fig_hmc_train_val_curve}, in the training phase, both accuracy and error seem to converge in the intended direction very smoothly. However, as in the case of the EDFX dataset, the model struggled to stabilize itself and exhibited a scattered performance over the entire epoch. The model tends to stabilize better as the number of batch sizes increases in the case of accuracy and error metrics, except for a large spike at approximately 50 epochs, starting from the batch size of 128.

The NCH dataset model performed better consistently from the beginning compared to the EDFX and HMC datasets. However, one interesting pattern to note here is that, unlike the cases of EDFX and HMC, where larger batch sizes positively impacted the model performance, the NCH dataset did not perform best for the largest batch size. For NCH, the model performed best with a batch size of 128, achieving an accuracy of 99.08\%, whereas with a batch size of 1024, it had an accuracy of 97.94\%. Although for the top-2 accuracy, the performance varied, in the case of the top-3 accuracy, the performance was consistent with an accuracy of 99.96\%. In the case of the error rate, it outperformed the other two datasets in terms of MAE and MSE, with the lowest values of 0.0136 and 0.1456, respectively. However, similar to the other two datasets, it has a smooth learning curve for both accuracy and loss values in training, and there are abrupt changes in the validation phases, slowly converging to their intended values (Supplementary Figure \ref{fig_nch_train_val_curve}).

Cohen's kappa is used to quantify the reliability of agreement between the two raters. If we consider the $\kappa$ values in Table \ref{tab_mdl_perf}, only four instances are below 0.90. According to Cohen's kappa interpretation \parencite{mchugh_interrater_2012}, a $\kappa$ value of 0.9 or greater implies an almost perfect level of agreement with approximately 82-100\% of reliable data. Therefore, according to the results, our model predicted almost perfectly, and if two different versions of our model predicted new random data, they would agree in most cases. Some exceptions were observed for several batch sizes in the EDFX and HMC datasets. For example, although the EDFX dataset has a strong level of agreement with 64-81\% reliable data, in the case of the first two batch sizes of the HMC dataset, there is none and a weak level of agreement. This indicates that if two instances of the former models attempt to predict a sleep stage, they will almost always contradict each other. The remaining model performance parameters, including precision, recall, and MF1, are listed in Table \ref{tab_mdl_perf}. In Supplementary Figures \ref{fig_mean_max_inter} and \ref{fig_mean_max_intra}, we show the mean and maximum performance of our model based on the inter-datasets and inter-batch size settings.

\begin{figure}
    \centering
    \includegraphics[width=0.6\textwidth]{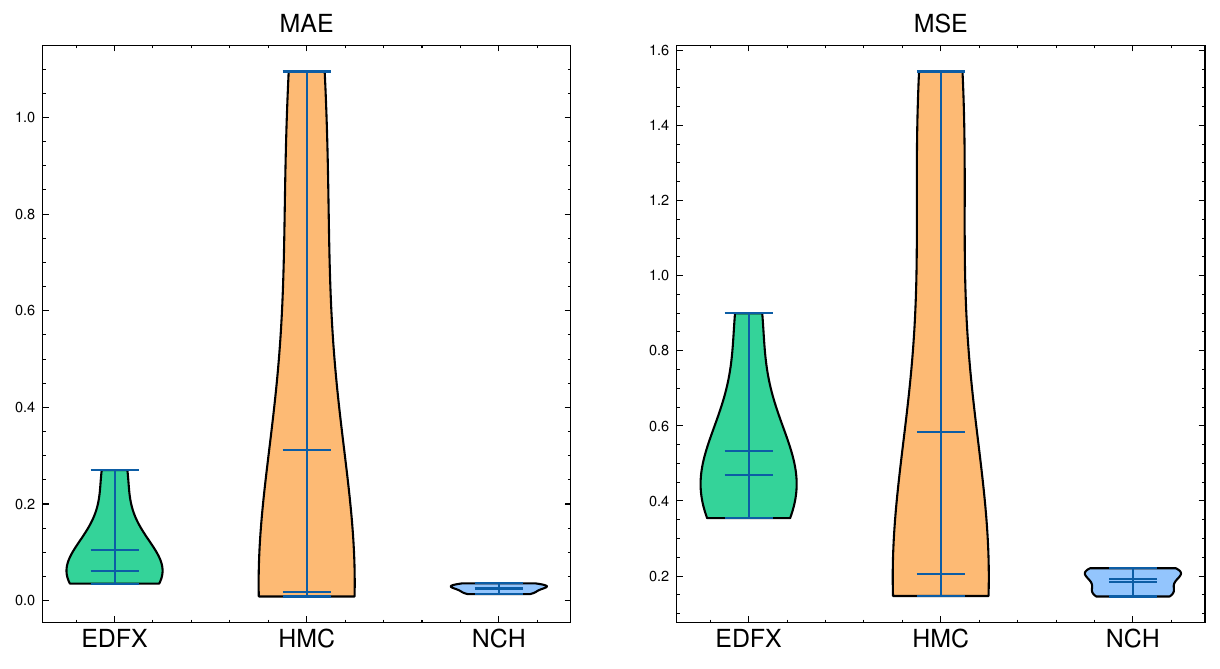}
    \caption{Representation of two error metrics for our model's performance across different batch sizes for the three datasets.}
    \label{fig_err_violin}
\end{figure}

Figure \ref{fig_err_violin} illustrates the overall loss of the model based on the three datasets for all batch sizes. It is quite noticeable that the NCH dataset performed very well, with very little loss compared to the EDFX dataset. However, HMC had the highest error percentage among the three datasets, with an approximate median value of 0.3. An interesting pattern to note is that, although the NCH had the lowest error, the plot width increased as the batch size increased, indicating that our model performed the worst, which was not the case for EDFX and HMC. Their loss decreased with increasing batch sizes.

Overall, the model was able to automatically classify sleep data most of the time. Although there were significant fluctuations during the earlier epochs, they became more stable later, and the validation performance was closely aligned with the training performance. The loss of the proposed function was consistently minimal, indicating good performance and the prevention of misdiagnosis.

\begin{table*}[!tb]
\centering
\resizebox{\textwidth}{!}{%
\begin{threeparttable}
\caption{Model's performance comparison with existing literature. }
\label{tab_perf_comp}
\begin{tabular}{@{}llllllllll@{}}
\toprule
Dataset & Ref. & Classifier & EEG Channel & \#Param & Recall & Precision & MF1 & Kappa $\kappa$ & Accuracy \\ \midrule
\multirow{21}{*}{EDF/EDFX} & \parencite{Li_Chen_Liu_Zhao_2023} & 4s-SleepGCN & Fpz-Cz & 2.50 M & 90.00 & 88.70 & 89.10 & 0.89 & 92.30 \\
 & \multirow{2}{*}{\parencite{Qu_Wang_Hong_Chi_Feng_Grunstein_Gordon_2020}} & \multirow{2}{*}{CNN+Attention} & Fpz-Cz & -- & -- & -- & 79.00 & 0.78 & 84.30 \\
 &  &  & Pz-Oz & -- & -- & -- & 74.10 & 0.74 & 80.70 \\
 & \parencite{Eldele_Ragab_Chen_Wu_Kwoh_Li_Guan_2023} & ADAST & Fpz-Cz & -- & -- & -- & 60.39 & -- & 74.00 \\
 & \parencite{Eldele_Chen_Liu_Wu_Kwoh_Li_Guan_2021} & AttnSleep & Fpz-Cz & -- & -- & -- & 78.10 & 0.79 & 84.40 \\
 & \parencite{Li_Chen_Cheng_2022} & ST-GCN & \begin{tabular}[c]{@{}l@{}}Fpz-Cz\\ Pz-Oz\end{tabular} & -- & 90.90 & 87.40 & 89.00 & 0.88 & 91.00 \\
 & \parencite{Zhu_Luo_Yu_2020} & CNN+Attention & Fpz-Cz & -- & -- & -- & 84.50 & -- & 93.70 \\
 & \parencite{Salman_Li_Oudah_Almaged_2023} & LS-SVM & \begin{tabular}[c]{@{}l@{}}Fpz-Cz\\ Pz-Oz\end{tabular} & -- & 96.50 & -- & -- & 0.87 & 97.40 \\
 & \multirow{2}{*}{\parencite{Abdulla_Diykh_Siuly_Ali_2023}} & \multirow{2}{*}{Genetic Algorithm} & Fpz-Cz & -- & 91.19 & -- & \textbf{--} & 0.92 & 92.41 \\
 &  &  & Pz-Oz & -- & 93.26 & -- & -- & 0.93 & 93.75 \\
 & \multirow{2}{*}{\parencite{Huang_Ren_Zhou_Yan_2022}} & \multirow{2}{*}{CSCNN-HMM} & Fpz-Cz & -- & -- & -- & -- & 0.79 & 84.60 \\
 &  &  & Pz-Oz & -- & -- & -- & -- & 0.76 & 82.30 \\
 & \parencite{Li_Qi_Ding_Zhao_Sang_Lee_2022} & EEGSNet & Fpz-Cz & -- & -- & -- & 77.26 & 0.77 & 83.02 \\
 & \multirow{2}{*}{\parencite{Zhou_Wang_Liu_Wu_Xu_Wang_Ye_Xia_Hu_Tian_2020}} & \multirow{2}{*}{RF-LGB} & Fpz-Cz & -- & -- & -- & 92.00\tnote{1} & 0.864 & 91.20 \\
 &  &  & Pz-Oz & -- & -- & -- & 92.00\tnote{1} & 0.872 & 91.80 \\
 & \parencite{Abdollahpour_Rezaii_Farzamnia_Saad_2020} & TLCNN-DF & \begin{tabular}[c]{@{}l@{}}Fpz-Cz\\ Pz-Oz\end{tabular} & -- & -- & 85.46\tnote{2} & 81.80\tnote{2} & 0.90 & 93.16 \\
 & \textbf{Ours} & \textbf{AttDiCNN} & \textbf{Fpz-Cz} & \textbf{1.41 M} & \textbf{98.90} & \textbf{98.63} & \textbf{98.77} & \textbf{0.98} & \textbf{98.56} \\ \midrule
\multirow{3}{*}{HMC} & \parencite{Zhou_Zhao_Wang_Jiang_Yu_Li_Li_Pan_2024} & Generative Model & -- & -- & -- & -- & 74.00\tnote{3} & -- & 77.70\tnote{3} \\
 & \parencite{Choi_Sung_2022} & DNN & -- & -- & -- & -- & -- & 0.70 & 77.00 \\
 & \textbf{Ours} & \textbf{AttDiCNN} & \textbf{C3-M2} & \textbf{1.41 M} & \textbf{99.62} & \textbf{99.60} & \textbf{99.61} & \textbf{0.99} & \textbf{99.66} \\ \midrule
\multirow{2}{*}{NCH} & \parencite{Lee_Saeed_2022} & Transformer Model & -- & -- & -- & -- & 70.50 & 0.71 & 78.20 \\
 & \textbf{Ours} & \textbf{AttDiCNN} & \textbf{C3-M2} & \textbf{1.41 M} & \textbf{99.14} & \textbf{99.39} & \textbf{99.26} & \textbf{0.99} & \textbf{99.08} \\ \bottomrule
\end{tabular}%
\begin{tablenotes}
    \item [1] No macro F1 was reported. Thus, a weighted F1 score was provided.
    \item [2] The values were derived by calculating the average of each individual sleep stage.
    \item [3] The values are based on the EOG data instead of the EEG data.
\end{tablenotes}
\end{threeparttable}
}
\end{table*}

\subsubsection{Inter Performance}

In this section, we evaluate the performance of our model against existing studies. Although we utilized three datasets to assess our model's performance, to the best of our knowledge, most existing studies primarily employ the EDF or EDFX datasets, with some studies using the HMC or NCH datasets. Most models in these studies were designed based on convolutional networks. For example, \textcite{Abdollahpour_Rezaii_Farzamnia_Saad_2020} introduced the TLCNN-DF framework, achieving an accuracy of 93.16\% and a kappa of 0.90, indicating a perfect level of agreement. Their framework integrates information from two data sources: EEG and electrooculography (EOG). However, obtaining both EEG and EOG data simultaneously from the same source may not always be practical. In addition, models that rely on transfer learning introduce computational complexities. \textcite{Li_Qi_Ding_Zhao_Sang_Lee_2022} developed EEGSNet, a multilayer CNN that automatically classifies sleep scores from EEG spectrograms. They also used a bidirectional long short-term memory model to capture transitional information from the extracted features. However, the retention of long-term memory throughout the model lifetime results in redundant information. Moreover, their model did not perform satisfactorily compared to the state-of-the-art results. In contrast, \textcite{Salman_Li_Oudah_Almaged_2023} proposed a cluster-based approach for robust sleep staging, outperforming numerous CNN-based methods, with an accuracy of 97.40\% and a recall of 96.50\%, which demonstrates strong model agreement. Furthermore, \textcite{Abdulla_Diykh_Siuly_Ali_2023} adopted an ensemble method that utilized genetic algorithms to classify sleep stages based on Fourier transformation. They evaluated the performance of their model on two distinct EEG channels, Fpz-Cz and Pz-Oz, achieving accuracies of 92.41\% and 93.75\%, respectively. Several authors have explored the incorporation of attention networks into deep learning layers to enhance the overall model performance. For instance, \textcite{Eldele_Chen_Liu_Wu_Kwoh_Li_Guan_2021} introduced AttnSleep, which employs a temporal context encoder with a multihead attention mechanism to classify sleep stages. To extract meaningful features more effectively, they proposed a multiresolution CNN and adaptive feature recalibration module. The CNN captures high- and low-frequency features, whereas the adaptive calibration module improves the quality of the features by modeling their interdependencies. Generally, they reported an overall accuracy of 84.40\% and an MF1 score of 78.10, indicating a balanced performance between precision and recall. Similarly, \parencite{Qu_Wang_Hong_Chi_Feng_Grunstein_Gordon_2020} and \parencite{Zhu_Luo_Yu_2020} both utilized attention mechanisms alongside convolutional networks but observed significant performance variations of approximately 9-13\%. \textcite{Zhu_Luo_Yu_2020} designed a CNN-based network to capture the characteristics of local EEG data, passing them to the attention network for better analysis of inter- and intra-epoch features. They achieved an overall accuracy of 93.70\% and an MF1 of 84.50 for the Fpz-Cz channel. In contrast, \textcite{Qu_Wang_Hong_Chi_Feng_Grunstein_Gordon_2020} achieved overall accuracies of 84.30\% and 80.70\% for the Fpz-Cz and Pz-Oz channels, respectively, by decomposing input EEG signals into different frequency bands and feeding them to CNN and MHA networks. As shown in Table \ref{tab_perf_comp}, our proposed model outperforms the state-of-the-art networks in all evaluation metrics, with significant differences and reduced computational complexity.

\subsection{Inspecting Performance Robustness}

\begin{figure}[!htb]
    \includegraphics[width=\columnwidth]{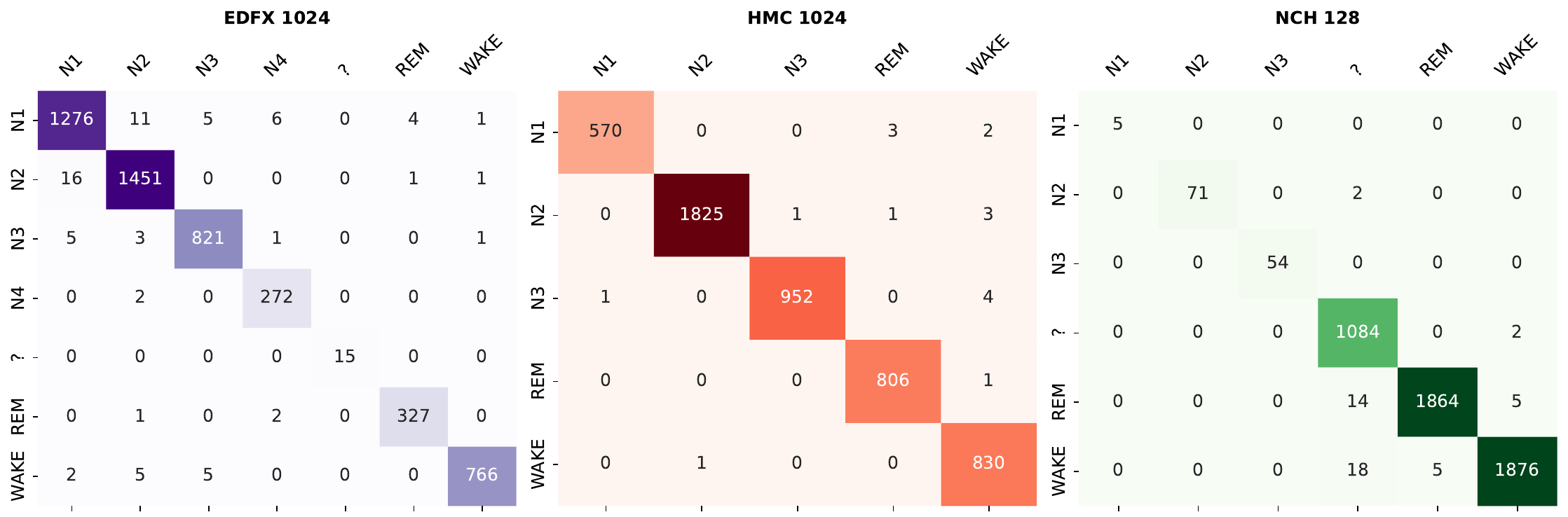}
    \caption{Confusion matrix displaying the best three models with batch sizes of 1024, 1024, and 128 for their respective datasets.}
    \label{fig_best_cf}
\end{figure}

\begin{figure*}[!htb]
    \includegraphics[width=\textwidth]{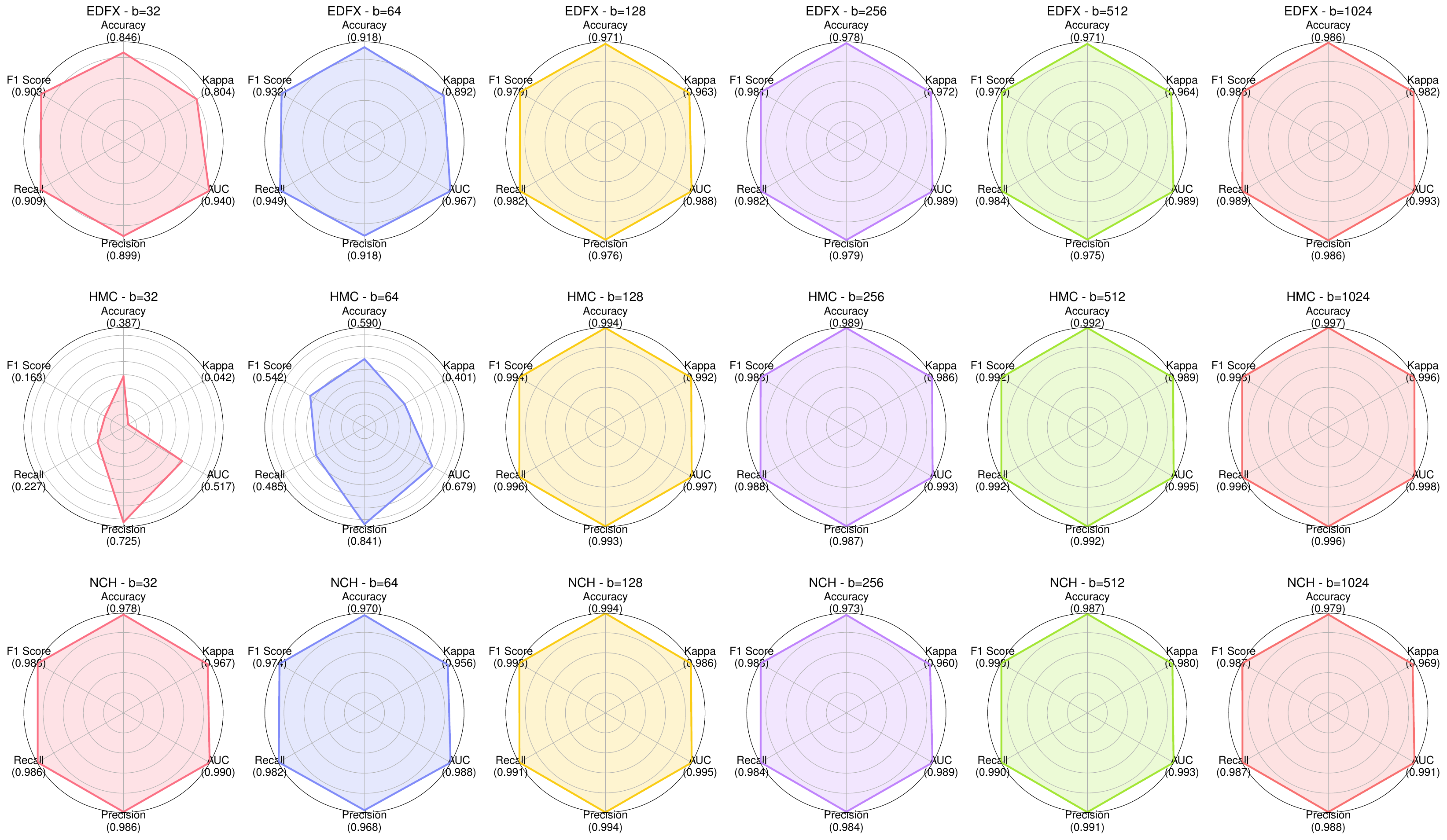}
    \caption{Reliability diagram of the model for the three datasets with batch sizes ranging from 32 to 1024, including six statistical metrics: accuracy, kappa, AUC, precision, recall, and F1 score.}
    \label{fig_mdl_perf_radar}
\end{figure*}

As shown in Figure \ref{fig_best_cf}, our proposed model predicted nearly all sleep stages accurately. For the EDFX dataset with a batch size of 1024, the N1 sleep stage exhibited the highest misclassification rate (27 instances). N1 was confused with the N2 stage in 11 out of 5000 samples. In contrast, the N2 class showed 18 misclassification cases, with 16 instances overlapping with the N1 stage. In addition, there were minor overlaps between the wake stage and N2 and N3 stages. The model distinctly identified the remaining stage. For the HMC dataset, also with a batch size of 1024, unlike the EDFX dataset, the model effectively resolved the confusion between the N1 and N2 stages, presenting zero overlap. However, some ambiguity was noted between the N2 and N3 sleep stages and the wake stage, with three and four overlapping instances, respectively. Furthermore, the model classified the other stages nearly perfectly. Similarly, for the NCH dataset with a batch size of 128, the model successfully ameliorated the N1 and N2 confusions, similar to the HMC dataset. However, a notable discrepancy was observed in the unknown (?) stage, which overlapped with the REM and wake stages in 14 and 18 instances.

In brief, out of 5000 test samples, the top-performing models successfully predicted 4928, 4983, and 4954 sleep stages for the EDFX-1024, HMC-1024, and NCH-128 datasets. Despite achieving accuracies of 98.56\%, 99.66\%, and 99.08\% for each dataset, an observable trend is that performance decreases as the total number of sleep stages increases. To address this issue, careful feature engineering of the datasets is necessary to ensure consistent signal properties, such as sampling frequency, channel data, and electrode usage, as well as other parameters. The performance of the remaining batch sizes for the three datasets, based on the confusion matrix, is shown in Supplementary Figure \ref{fig_rest_cf}. However, disregarding the error rates of 1.44\%, 0.34\%, and 0.92\% for datasets with seven, five, and six sleep stages, respectively, the model's consistent performance across all datasets suggests its viability as an automatic sleep staging tool.

\begin{figure*}[!htb]
    \includegraphics[width=0.95\textwidth]{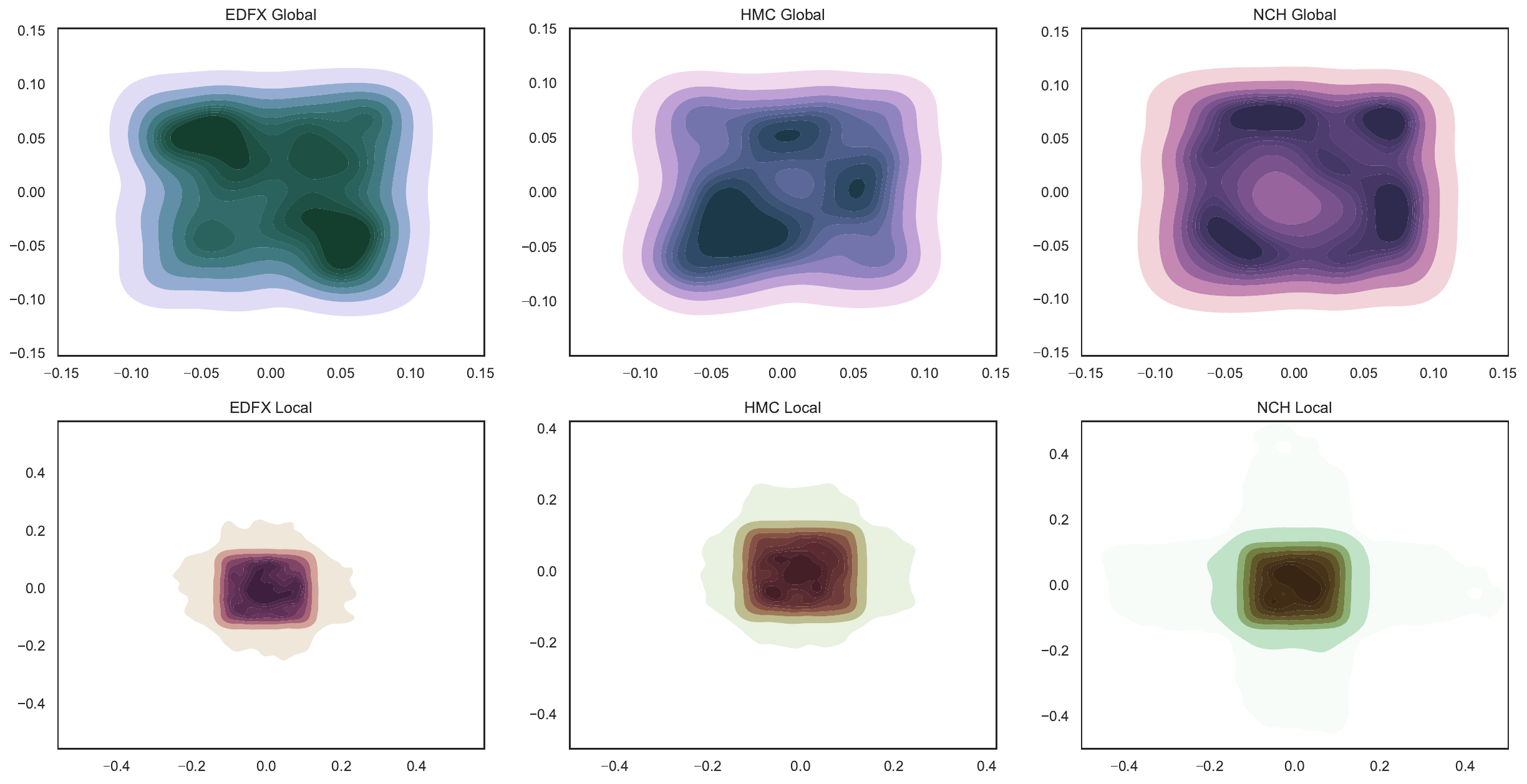}
    \caption{Kernel weight distribution comparison between the G2A module and the LSFE module to assess computational overhead.}
    \label{fig_weight_dist}
\end{figure*}

To further evaluate the robustness of the model, we illustrated a reliability diagram, as shown in Figure \ref{fig_mdl_perf_radar}. The underlying idea of the reliability diagram is that a larger diagram area implies a more reliable model. The six parameters included in this diagram were accuracy, kappa, AUC, precision, recall, and F1 score. Because higher values of these parameters indicate better performance in their respective criteria, larger values result in a greater area, indicating a superior model. This setup allowed us to visualize the performance across datasets of the same batch size and across different batch sizes of the same datasets. In the EDFX dataset, the model generally performed consistently, accurately classifying the sleep stages, except for the model with a batch size of 32, which covered a smaller area in the performance metrics. The NCH model displayed an almost perfect area under the curve across all batch sizes. However, in the HMC dataset, a noticeable change was observed in the covered area for the first two batch sizes. For these cases, the precision was higher than that of the other parameters, suggesting that when the models classify EEG data into a specific sleep stage, there is a high likelihood that the data belong to that stage. Nevertheless, other parameters, such as accuracy, recall, and AUC, showed poor performance relative to the neighboring batch sizes, indicating a lower probability of correctly classifying the sleep stages. Among all parameters, the kappa value was the lowest for these two batch sizes, indicating zero and weak agreement levels for batch sizes of 32 and 64, respectively. Furthermore, for the same batch size, the performance was fairly consistent across all datasets, except for the aforementioned batch sizes. Overall, by examining the areas of the reliability plots, it is evident that there is a trend of improved performance with increasing batch sizes.

\subsection{Attention Network Impact on Kernel Weight Distribution}

As stated previously, our proposed model incorporates a G2A module to extract the most pertinent information while alleviating the computational load. We examined the kernel weight distribution of specific modules in our model, as shown in Figure \ref{fig_weight_dist}. For the EDFX dataset, regarding the LSFE or local pattern observer, the weights were distributed throughout the range of $[-0.4, 0.4]$ in both directions. In addition, it highlights two main areas of interest: the less influential region (LIR) and the influential region (IR). In the LIR, numerous scattered weight distributions form an amorphous colony on which the LIRs are established. In the IR, a densely populated weight region with a rectangular configuration exists. Comparatively, the G2A module's weight distribution for the EDFX dataset also has a rectangular shape, similar to the IR of the LSFE module. Although the IRs of the LSFE and G2A regions share a similar shape, they differ in weight amplitude and their effect on computational overhead. In the IR of LSFE, there is a large dark area, whereas for G2A, there are two smaller regions with lower amplitudes. The segmentation of their weight regions eliminates some redundant weights, corresponding to irrelevant long-retention information of local features. This pattern was consistent with the other two datasets. The LSFE region spanned approximately $[-0.5, 0.5]$, whereas the G2A region spanned approximately $[-0.1, 0.1]$. The LSFE region has several dense areas that capture both relevant and irrelevant data, whereas the G2A regions consist of multiple smaller dense areas that contain the most relevant information and exclude redundant data. This exclusion significantly influenced the computational complexity of the model, reducing the total number of parameters to 1.41 M, compared to the model of \textcite{Li_Chen_Liu_Zhao_2023}, which had 2.50 M parameters and lower performance results, as detailed in Table \ref{tab_perf_comp}.

\section{Conclusion}

In this study, we proposed an automated sleep stage classifier called AttDiCNN, which leverages an attentive dilated receptive field. Force-directed layouts were generated from visibility graphs to extract the most significant features from the EEG signals. The architecture is divided into three modules: LSFE, S2TLR, and G2A. LSFE captures localized spatial features, which are then fed into the S2TLR network to capture long-term retention information via a self-attention mechanism. Finally, G2A unifies the local and global features by averaging the extracted features. We used three datasets, EDFX, HMC, and NCH, to evaluate the performance of our model. The model achieved state-of-the-art performance, with accuracies of 98.56\%, 99.66\%, and 99.08\% for the respective datasets. We believe that our approach has the potential to be a role model in the domain of sleep stage classification owing to its unique feature extraction methods and the use of advanced neural network techniques. In addition, owing to its lightweight nature, we hope that it can be applied in clinical settings to support healthcare professionals. However, we believe that future research can explore other aspects of the problem. For example, although we tested our model on multiple datasets, it is essential to generalize the model by training it on a merged dataset comprising various smaller datasets from different sources. The development of a univariate feature extraction method that can reliably predict new, unseen data would further increase its potential. It is also important for the model to explain its reasoning behind the predictions to support healthcare professionals in making the right decisions in clinical settings.

\section*{Acknowledgement}

The authors gratefully acknowledge the BIOSE Research Center at BRAC University for their valuable support in conducting this study. We also acknowledge the financial support provided by the BSRM School of Engineering at BRAC University.

\printbibliography

\clearpage

\renewcommand{\thefigure}{S\arabic{figure}}
\renewcommand{\thetable}{S\arabic{table}}
\setcounter{figure}{0}
\setcounter{table}{0}

\begin{center}
{\LARGE\bfseries Supplementary Materials\par}
\vspace{1ex}
\end{center}

\begin{figure*}[!htb]
    \includegraphics[width=\textwidth]{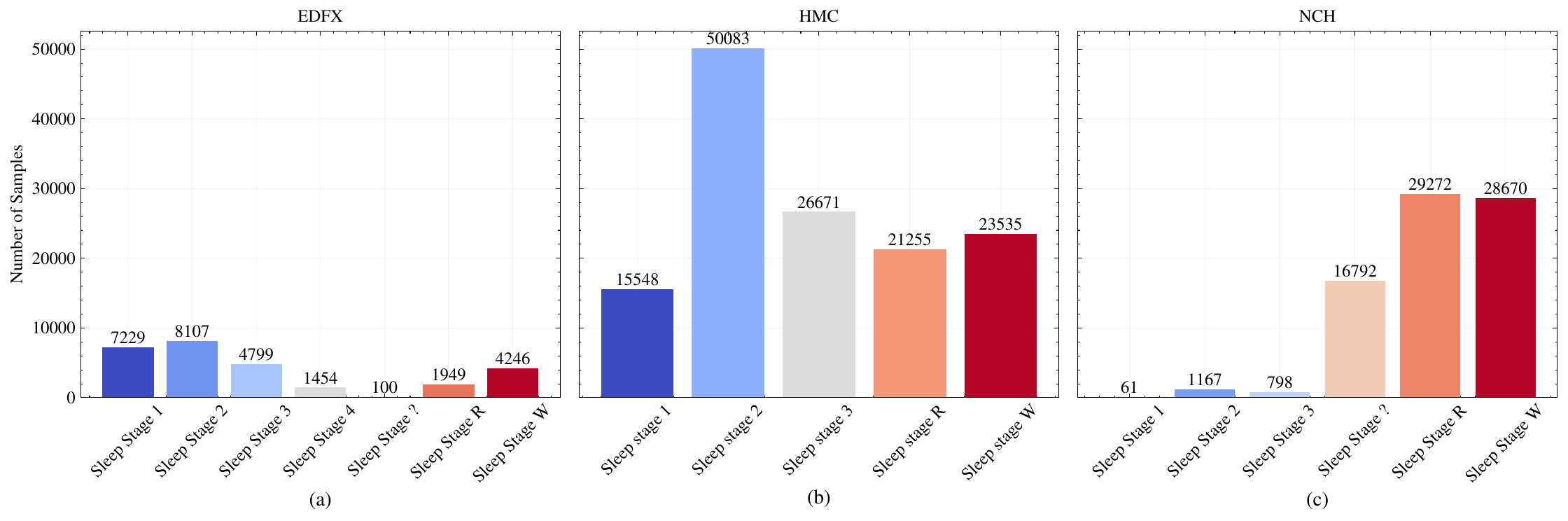}
    \caption{Class distribution of the three datasets: EDFX, HMC, and NCH. The datasets are heavily biased towards some particular classes (sleep stages). Sleep stages 1 and 2 dominate the EDFX and HMC datasets, whereas sleep stages R and W are more prevalent in the NCH dataset. Some classes, like sleep stage 4, and sleep stage ``?'' (no score) are exclusive to one or more particular datasets.}
    \label{fig_class_dist}
\end{figure*}

\begin{figure*}[!htb]
    \includegraphics[width=\textwidth]{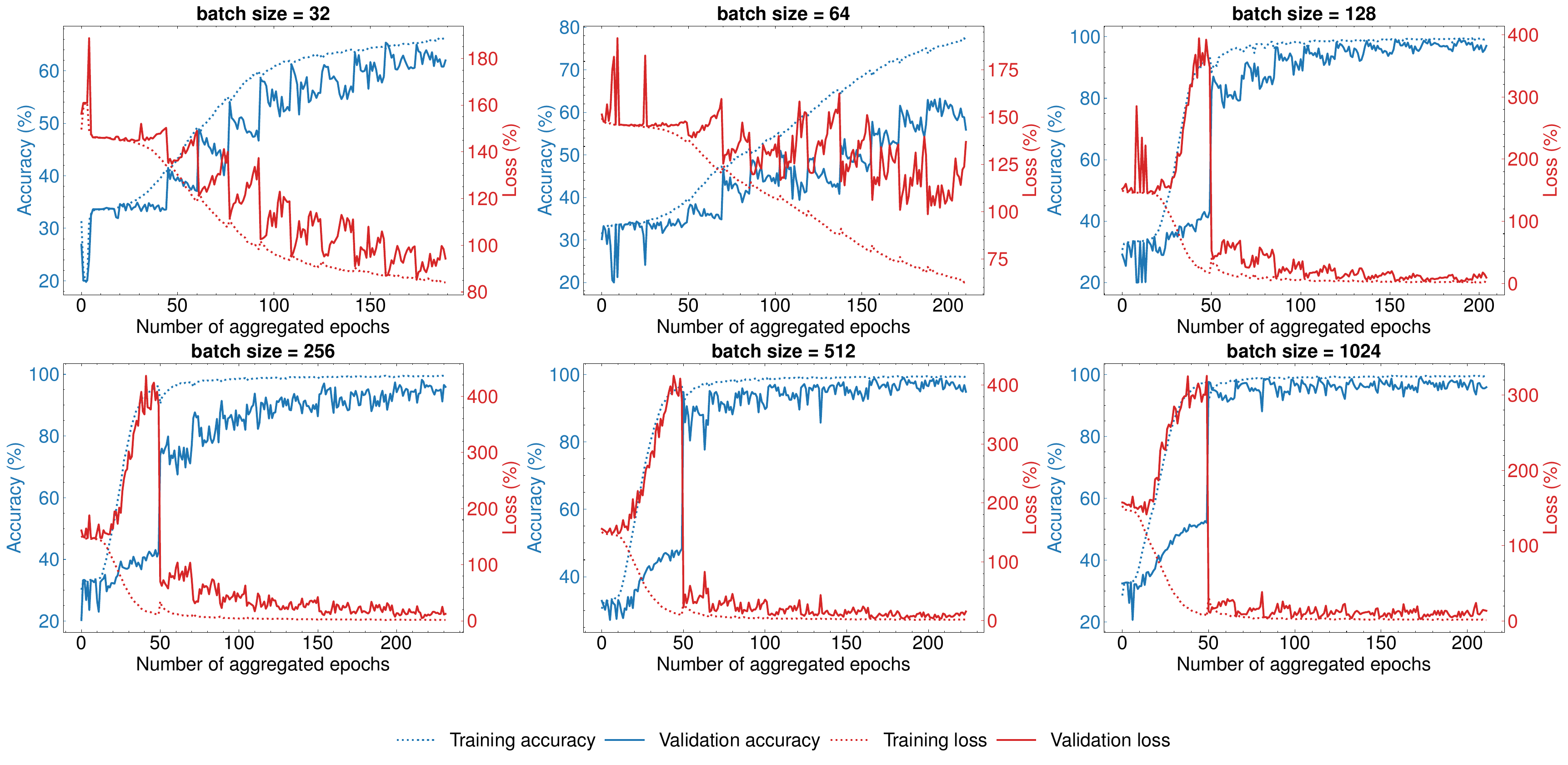}
    \caption{Performance metrics (accuracy and loss) plotted over epochs during training and validation for the HMC dataset, using batch sizes of 32, 64, 128, 256, 512, and 1024.}
    \label{fig_hmc_train_val_curve}
\end{figure*}

\begin{figure*}[!htb]
    \includegraphics[width=\textwidth]{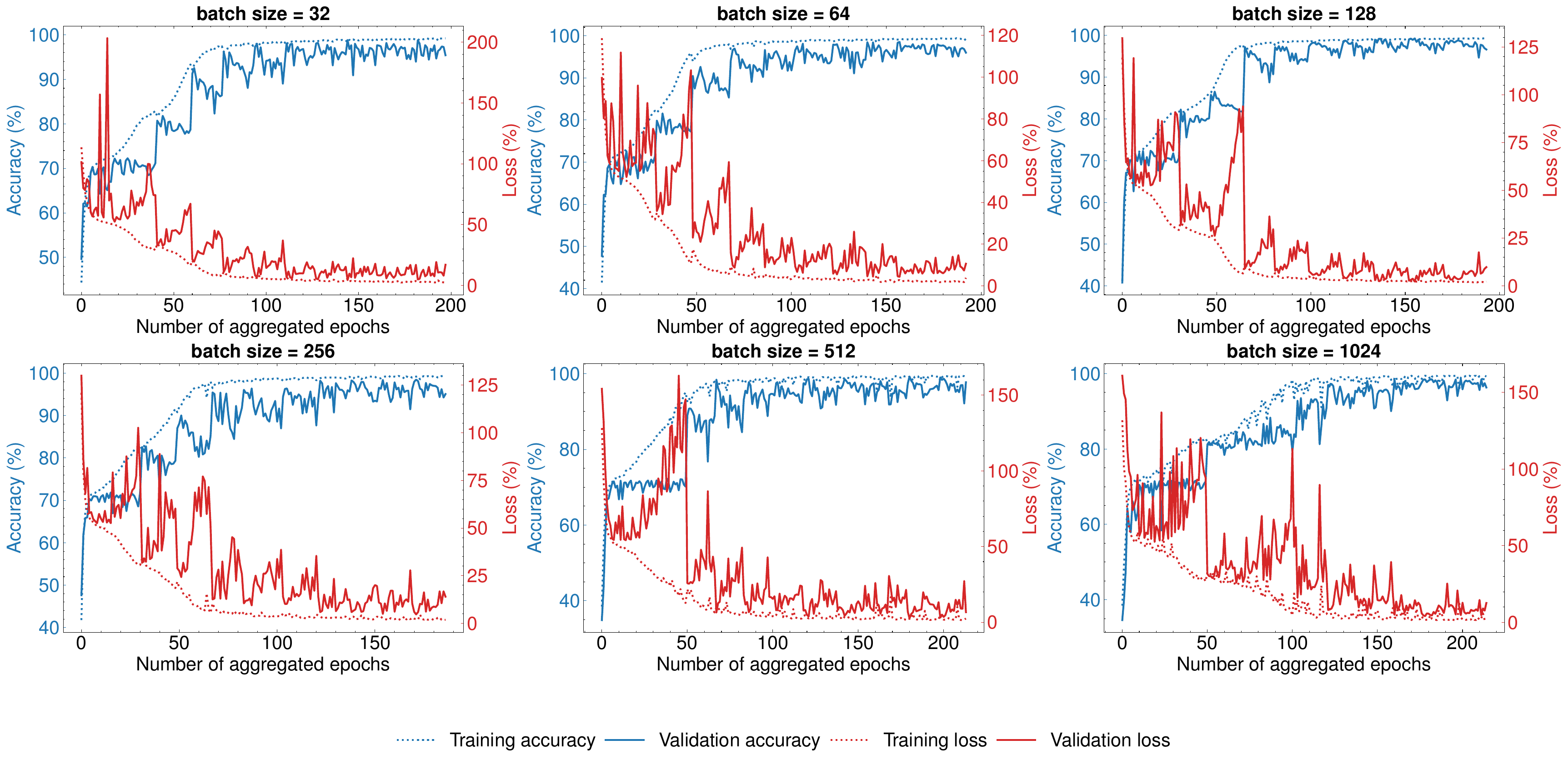}
    \caption{Performance metrics (accuracy and loss) plotted over epochs during training and validation for the NCH dataset, using batch sizes of 32, 64, 128, 256, 512, and 1024.}
    \label{fig_nch_train_val_curve}
\end{figure*}

\begin{figure*}[!htb]
    \includegraphics[width=\textwidth]{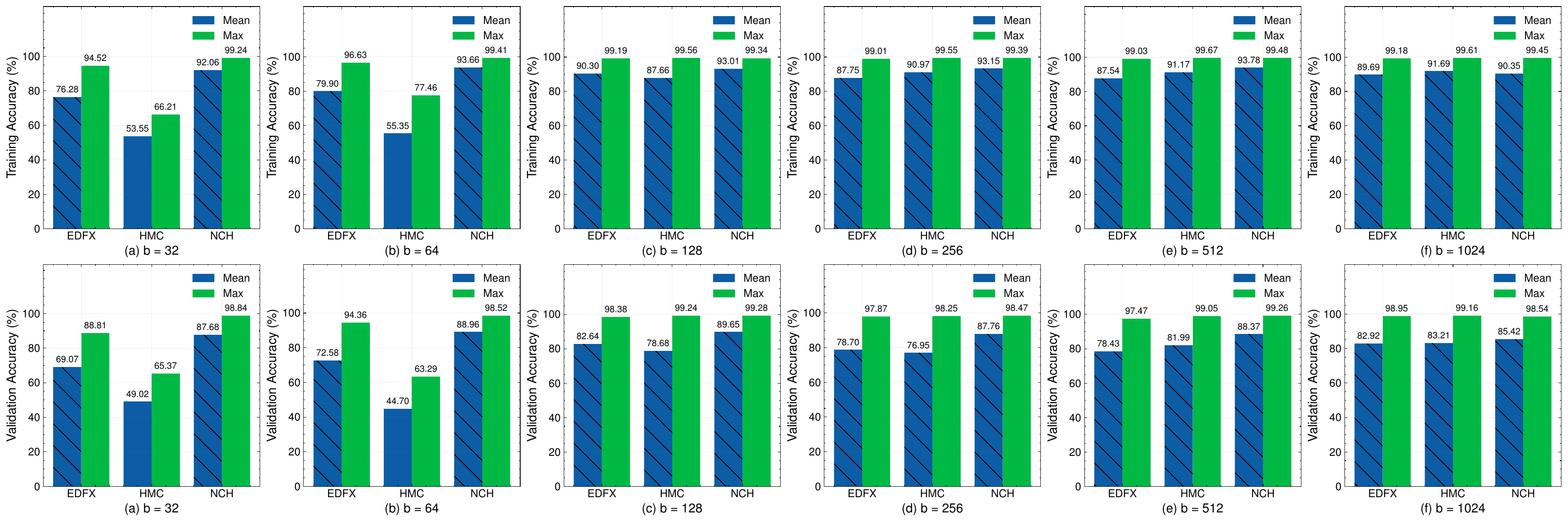}
    \caption{Analysis of maximum and average performance across three datasets during both training and validation phases, using different batch sizes.}
    \label{fig_mean_max_inter}
\end{figure*}

\begin{figure*}[!htb]
    \centering
    \includegraphics[width=\textwidth]{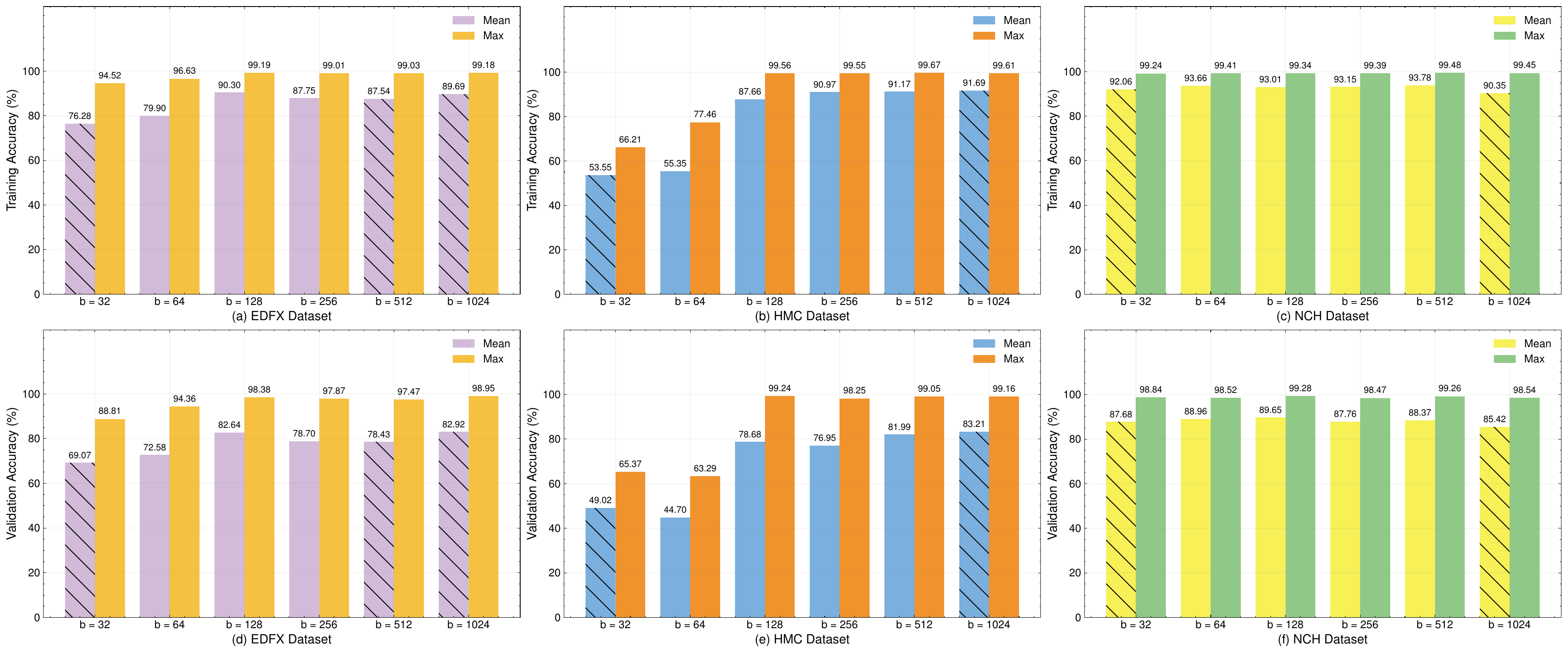}
    \caption{Comparison of maximum and mean performance among the three datasets' batch sizes during training and validation periods.}
    \label{fig_mean_max_intra}
\end{figure*}

\begin{figure*}[!htb]
    \includegraphics[width=\textwidth]{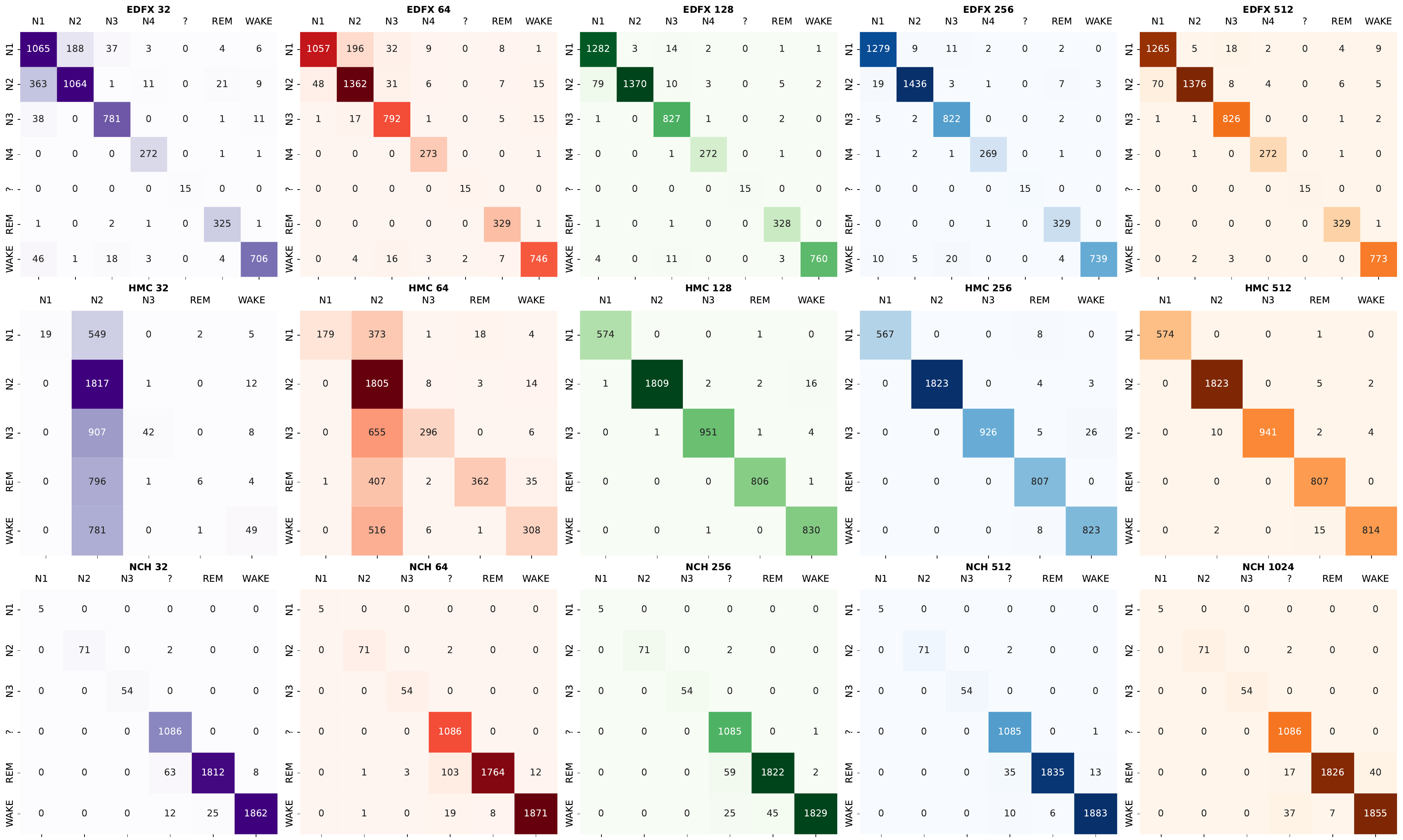}
    \caption{Confusion matrix illustrating the performance for non-optimal batch sizes across three different datasets.}
    \label{fig_rest_cf}
\end{figure*}

\end{document}